\documentclass[prx,aps,twocolumn,amsmath,amssymb]{revtex4-1}

\usepackage[margin=1in]{geometry}

\usepackage{graphicx}
\usepackage{cleveref}

\crefformat{equation}{#2#1#3}
\crefrangeformat{equation}{#3#1#4--#5#2#6}

\usepackage{amsmath}
\usepackage{amssymb}
\usepackage{mathrsfs}
\usepackage{physics}

\begin{document}

\title{On the nature of the glass transition in atomistic models of glass formers}

\author{Alexander Hudson}
\thanks{We are deeply indebted to David Chandler, without whom this work would not have been possible. David was intimately involved with this work from the beginning and we regret that it could not be completed before he passed away. We wish to emphasize that the opinions and perspective expressed in this paper belong solely to the listed authors, AH and KKM. Although David's intellectual contributions to this work warrant his inclusion as an author on this paper, we felt it inappropriate to add his name without his permission. If not for David's untimely passing, he too would have contributed to the writing of the manuscript and likely suggested significant revisions to meet his exacting standards.}
\author{Kranthi K.\ Mandadapu}
\thanks{We are deeply indebted to David Chandler, without whom this work would not have been possible. David was intimately involved with this work from the beginning and we regret that it could not be completed before he passed away. We wish to emphasize that the opinions and perspective expressed in this paper belong solely to the listed authors, AH and KKM. Although David's intellectual contributions to this work warrant his inclusion as an author on this paper, we felt it inappropriate to add his name without his permission. If not for David's untimely passing, he too would have contributed to the writing of the manuscript and likely suggested significant revisions to meet his exacting standards.}

\affiliation{Department of Chemical and Biomolecular Engineering, University of California, Berkeley,}
\affiliation{Chemical Sciences Division, Lawrence Berkeley National Laboratory}

\begin{abstract}
We study the nature of the glass transition by cooling model atomistic glass formers at constant rate from a temperature above the onset of glassy dynamics to $T=0$. Motivated by the East model, a kinetically constrained lattice model with hierarchical relaxation, we make several predictions about the behavior of the supercooled liquid as it passes through the glass transition. We then compare those predictions to the results of our atomistic simulations. Consistent with our predictions, our results show that the relaxation time $\tau$ of the material undergoes a crossover from super-Arrhenius to Arrhenius behavior at a cooling-rate-dependent glass transition temperature $T_\text{g}$. The slope of $\ln\tau$ with respect to inverse temperature exhibits a peak near $T_\text{g}$ that grows more pronounced with slower cooling, matching our expectations qualitatively. Additionally, the limiting value of this slope at low temperature shows remarkable quantitative agreement with our predictions. Our results also show that the rate of short-time particle displacements deviates from the equilibrium linear scaling around $T_\text{g}$, asymptotically approaching a different linear scaling. To our surprise, these short-time displacements, the dynamic indicators of the underlying excitations responsible for structural relaxation, show no spatial correlations beyond a few particle diameters, both above and below $T_\text{g}$. This final result is contrary to our expectation, based on previous results for East model glasses formed by cooling, that inter-excitation correlations should emerge as the liquid vitrifies.
\end{abstract}

\maketitle

\section{Introduction}

A glass may be understood as a liquid whose structural relaxation time, $\tau$, the timescale over which its constituent particles can rearrange, exceeds the accessible timescale of observation. %% Perhaps cite a review here?
On timescales short compared to the structural relaxation time, i.e., $t\ll\tau$, the liquid does not flow, instead demonstrating the rigidity of a solid despite lacking obvious structural ordering. Vitrification, the process by which a flowing liquid transforms into a rigid glass, can be attributed to the explosive growth in the relaxation time that accompanies decreasing temperature in supercooled liquids. Specifically, it is well known that the relaxation time of a liquid exhibits super-Arrhenius growth %% Citation for super-Arrhenius growth
at temperatures below a material-dependent onset temperature $T_\text{o}$ \cite{corresponding-states-i,corresponding-states-ii}. For $T<T_\text{o}$, $\tau(T)$ grows super-exponentially with the inverse temperature. For historical overviews of the field, covering both theory and experiment, see references \cite{ediger-angell-nagel,angell-review-1996,angell-review-2000,debenedetti-stillinger-review,biroli-garrahan-review}.

Although there is disagreement about the precise form of $\tau(T)$, %% Citation
in this paper we take the perspective of the East model \cite{east-1991}, a kinetically constrained lattice model of glass formers, for which the relaxation time has been shown theoretically and numerically to obey parabolic scaling, \cite{sollich-evans-1999,sollich-evans-2003,dayton-east-model,aldous-east-model-2002,chleboun-east-model-2013}
\begin{equation}\label{eq:parabolic_law}
  \ln\tau(T) \sim J^2\left(\frac{1}{T}-\frac{1}{T_\text{o}}\right)^2\quad \text{for } T\ll T_\text{o},
\end{equation}
where $J$ is a material property and is related to the free-energetic cost of creating an excitation --- a localized region of space where particles can undergo nontrivial short-time displacements --- in the supercooled liquid \cite{keys-prx-2013}. %% Citation (probably Keys PRX)
Equation \ref{eq:parabolic_law} is noteworthy in that it remains finite at all nonzero temperatures, in contrast to other functional forms proposed for $\tau(T)$, such as the Vogel-Fulcher-Tammann equation common in the literature \cite{ediger-angell-nagel}, that diverge at finite temperature. %% Citation for other forms diverging at finite temperature

%% Probably just explicitly point interested readers to a set of reviews and leave it at that.

The origin of the parabolic scaling in equation \ref{eq:parabolic_law} is a relaxation barrier that grows linearly with inverse temperature, $E(T)\propto 1/T$. In the East model, this linear scaling is a consequence of two factors. The first is hierarchical dynamics, which dictates that the free-energetic barrier to relaxing an inter-excitation domain of length $\ell$ grows logarithmically with $\ell$, i.e., %% Cite Sollich and Evans [Comment: Did this at the end of the sentence below]
\begin{equation}\label{eq:hierarchical}
  E_\ell = \gamma J_0\ln\ell,
\end{equation}
where $J_0$ is the energy associated with an excitation and $\gamma$ measures the entropy of relaxation pathways \cite{sollich-evans-1999,sollich-evans-2003,chleboun-east-model-2013}. %% Citation for entropy of relaxation pathways. (Note that many papers try to work out gamma analytically.)
The second factor is the growing lengthscale between excitations as temperature is lowered. Because excitations cost energy, their concentration will decrease with temperature according to $c(T)\approx\exp\left(-J_0/T\right)$, so that the mean inter-excitation spacing $\ell=1/c$ will obey
\begin{equation}\label{eq:spacing}
  \ell(T) \approx e^{J_0/T}.
\end{equation}
Combining equations \ref{eq:hierarchical} and \ref{eq:spacing}, we find that $E(T)\approx\gamma J_0^2/T$. The relaxation time is then given by
\begin{equation}\label{eq:parabolic_law_derived}
  \tau(T) = \tau_0 e^{E(T)/T} \approx \tau_0 e^{\gamma J_0^2/T^2}.
\end{equation} %% Check S&E to see how much is worked out there. Also cite DL.
This form for the relaxation time of the East model was first derived by Sollich and Evans in reference \cite{sollich-evans-1999} and later confirmed analytically \cite{aldous-east-model-2002,chleboun-east-model-2013}. %% Do we need this here? I needed a place to cite Sollich-Evans for the final step of the derivation.
Comparing equation \ref{eq:parabolic_law_derived} to equation \ref{eq:parabolic_law}, we see that the curvature parameter in the parabolic law is related to the excitation energy scale via $J^2=\gamma J_0^2$.

Regardless of the exact form of $\tau(T)$, the super-Arrhenius growth of the relaxation time ensures that a modest lowering of temperature can transform an ergodic liquid into a non-ergodic glass, the latter unable to relax on the finite timescale of interest. Although it is possible, in principle, to prepare an equilibrium liquid at any temperature $T>0$, in practice the finite timescale implicit in any preparation process, whether performed in the laboratory or on a computer, will be dwarfed by the relaxation time of the liquid at sufficiently low temperatures.

To make the preceding discussion concrete, we limit our consideration to protocols in which a bulk liquid is in contact with a heat bath whose temperature follows a schedule $T(t)$. Although $T(t)$ could be a complicated function, in this paper we focus exclusively on constant-rate cooling protocols, such that $T(t)$ decreases linearly from an initial temperature $T_\text{i}$ to a final temperature $T_\text{f}$ in a time $t_\text{c}$. Assuming $T_\text{i}$ is a high temperature, say $T_\text{i}\gtrsim T_\text{o}$, the protocol is entirely characterized by just two parameters: the final temperature $T_\text{f}$ and the cooling rate $\nu_\text{c}\equiv\left|dT(t)/dt\right|=\left(T_\text{i}-T_\text{f}\right)/t_\text{c}$.

The inverse cooling rate, $\nu_\text{c}^{-1}$, is the timescale over which the protocol produces significant changes in temperature. If this timescale is long compared to changes in the relaxation time of the liquid as temperature is lowered, then the liquid will remain in equilibrium. Mathematically, we can express this condition as $\nu_\text{c}^{-1}\gg\left|d\tau/dT\right|$. If the opposite is true, $\nu_\text{c}^{-1}\ll\left|d\tau/dT\right|$, then the timescale set by the protocol is short compared to changes in the relaxation time and the liquid will fall out of equilibrium.

If we choose the final temperature of the protocol to be $T_\text{f}=0$, then there will be a crossover at an intermediate temperature $T_\text{g}$ where the two timescales are equal,
\begin{equation}\label{eq:define_Tg}
  \nu_\text{c}^{-1} = \left|\frac{d\tau(T_\text{g})}{dT}\right|.
\end{equation} %% See if this definition is used elsewhere
For $T>T_\text{g}$, the liquid remains in equilibrium; for $T<T_\text{g}$, the liquid is out of equilibrium. For $T\ll T_\text{g}$, the liquid has fully transformed into a glass. Thus, equation \ref{eq:define_Tg} formally defines the glass transition temperature $T_\text{g}$ \cite{Limmer-PNAS-2014,Limmer-JCP-2014}. Although other definitions are possible, %% Cite A. Angell
equation \ref{eq:define_Tg} clearly expresses the concept of the glass transition as a dynamic process in which an equilibrium supercooled liquid gradually transforms into a non-ergodic material. Equation \ref{eq:define_Tg} also makes explicit the protocol dependence of the glass transition. Note that if $\tau(T)$ is non-singular for all $T>0$, as it is for the parabolic law, then in the limit of infinitely slow cooling, $\nu_\text{c}\rightarrow 0$, equation \ref{eq:define_Tg} implies there is no finite-temperature glass transition.

In this paper, we subject two atomistic models of glass formers commonly used in simulation studies of glassy dynamics, the Wahnstr\"{o}m and Kob-Andersen models \cite{Wahnstrom-1991,Kob-Andersen-1995}, %% Cite models
to constant-rate cooling as described above. Using equation \cref{eq:define_Tg} and the East model relations, equations \cref{eq:parabolic_law,eq:hierarchical,eq:spacing,eq:parabolic_law_derived}, we make predictions about the glass transition and compare those predictions to the results of our atomistic cooling protocols. %% Perhaps mention L. Berthier's models here (though they are mentioned later)

Consistent with our prior expectations, we find a crossover in the relaxation time from the parabolic scaling of equation \ref{eq:parabolic_law} to linear (Arrhenius) scaling. This crossover is characterized by a peak in the slope of $\ln\tau(T)$ versus $1/T$, which grows more pronounced with slower cooling. This behavior agrees qualitatively with our predictions, which are based on approximate but intuitive arguments. Specifically, the slope exhibits its peak near $T_\text{g}$ given by equation \ref{eq:define_Tg}. Furthermore, our prediction for the slope at $T\ll T_\text{g}$, which we obtain by evaluating the energy barrier in equation \ref{eq:hierarchical} at $T_\text{g}$, agrees quantitatively with the simulation results for the Kob-Andersen model.

In addition to the super-Arrhenius-to-Arrhenius crossover, we also observe a crossover in the rate at which particles displace over short times $t\ll\tau$, from the equilibrium linear scaling established in reference \cite{keys-prx-2013} to a different linear scaling below $T_\text{g}$. Because the rate of short-time displacements is closely related to the concentration of excitations in the East model, we interpret this result as indicating that the number of excitations in the liquid approaches a constant, limiting value for $T\ll T_\text{g}$, with the residual slope corresponding to a temperature-independent free-energetic barrier to short-time dynamics within an excitation.

To our surprise, the spatial distribution of short-time particle displacements remains ideal-gas-like in the atomistic models, even for $T<T_\text{g}$. To the extent that these short-time displacements are faithful indicators of underlying excitations (which are otherwise undetectable), this contradicts East model results showing pronounced inter-excitation correlations emerging below $T_\text{g}$ \cite{keys-prx-2013,keys-pnas-2013,keys-pre-2015}. %% Cite Aaron's two papers
These correlations are a natural consequence of the East model's hierarchical dynamics, and their absence in the spatial distribution of particle displacements below $T_\text{g}$ is puzzling.

\section{Models and Methods}

\begin{table*}[t]
  \centering
  \small
  \begin{tabular}{l||c|c|ccc|ccc|cc||c}
    Model & $\rho$ & $f_A$ & $\sigma_{AA}$ & $\sigma_{AB}$ & $\sigma_{BB}$ & $\epsilon_{AA}$ & $\epsilon_{AB}$ & $\epsilon_{BB}$ & $m_A$ & $m_B$ & $T_\text{o}$ \\
    \hline
    W  & 1.296 & 0.5 & 1 & $11/12$ & $5/6$  & 1 & 1   & 1   & 2 & 1 & 0.82 \\
    KA & 1.2   & 0.8 & 1 & $0.8$   & $0.88$ & 1 & 1.5 & 0.5 & 1 & 1 & 0.72 \\
  \end{tabular}
  \caption{Parameters of the models considered in this work. The parameters listed between the double column bars fully define the models. Definitions of these parameters are as follows: $\rho$ is the number density $N/V$; $f_A$ is the fraction of particles that are type $A$; $\sigma_{ij}$ and $\epsilon_{ij}$ are the Lennard-Jones parameters for $ij$ pairs; and $m_i$ is the mass of type-$i$ particles. The last parameter in the table, $T_\text{o}$, is onset temperature as determined by fitting a parabolic law to equilibrium relaxation time data. At $T<T_\text{o}$, the models exhibit glassy dynamics.}
  \label{tab:params}
\end{table*}

\subsection{Models}

We consider binary mixtures of Lennard-Jones particles, which are governed by pairwise potentials of the form
\begin{equation}
  u_{\alpha\beta}(r) = -4\epsilon_{\alpha\beta}\left[\left(\frac{\sigma_{\alpha\beta}}{r}\right)^{6} - \left(\frac{\sigma_{\alpha\beta}}{r}\right)^{12}\right],
\end{equation}
where subscripts $\alpha,\beta$ denote the particle type, $A$ or $B$. The models chosen for this study are characterized by the Lennard-Jones parameters $\epsilon_{\alpha\beta}$ and $\sigma_{\alpha\beta}$, the dimension $d$, and the mole fraction $f_A$ of $A$-type particles. Additionally, for each model we fix the total density, $\rho\equiv N/V$, choosing values common in the literature. With the model and density specified, the behavior of the supercooled liquid --- characterized by parameters such as the onset temperature $T_\text{o}$ and energy scale $J$ --- is determined \cite{corresponding-states-i,corresponding-states-ii,keys-prx-2013}.

Binary mixtures are common choices in studies of glass-forming liquids due to their resistance to crystallization, which enables study of the liquid into the supercooled regime. We consider two binary mixture models in this paper: the Wahnstr\"{o}m model, a 50-50 mixture, and the widely studied Kob-Andersen model, an 80-20 mixture with non-additive parameters \cite{Wahnstrom-1991, Kob-Andersen-1995}. % Both models display glassy dynamics at temperatures below their respective onset temperatures and avoid crystallization to sufficiently cold temperatures that they are suitable for the study of glassy dynamics \cite{keys-prx-2013, Wahnstrom-1991, Kob-Andersen-1995}.
Table \ref{tab:params} lists the parameters for these models.

At sufficiently low temperatures, both models will succumb to crystallization. In our equilibrium simulations, we observe the Wahnstr\"{o}m model crystallizing at temperatures satisfying $T^{-1}\equiv\beta\gtrsim 1.7$ \cite{Pedersen-JCP-2009,Pedersen-PRL-2010,Pedersen-arxiv-2007} %% Might want to double check that these are the correct references
and the Kob-Andersen model at temperatures $\beta\gtrsim 2.3$. Given our estimates of $T_\text{o}$ for these two models, the temperature ranges over which we can safely study the supercooled liquid are $1.2\lesssim\beta\lesssim 1.7$ for the Wahnstr\"{o}m model and $1.4\lesssim\beta\lesssim 2.3$ for the Kob-Andersen model. The relatively narrow temperature range over which the Wahnstr\"{o}m model is a valid model glass former makes it a somewhat poor choice for studies of the glass transition, as the relatively high melting point imposes strict limits on the cooling rates that may be used. Nevertheless, we include results for cooling rates that are relatively fast, such that crystallization is avoided. The Kob-Andersen model, by contrast, has a much lower melting point and is consequently better suited for studies of the glass transition. Recently developed models characterized by continuous polydispersity offer even greater resistance to crystallization than the binary Kob-Andersen model \cite{Berthier-2017-models} and would make excellent choices for future studies extending the work in this paper, particularly if slower cooling rates could be accessed. %% I wanted to mention here that the game-changing speed-up offered by the algorithm/models in that paper unfortunately doesn't extend to what we're doing here, because we're preparing non-equilibrium states. The speed-up in that paper is only applicable to the preparation of equilibrium states, and the cooling protocols we use to prepare our non-equilibrium states must use the "real" dynamics of the system. However, it seemed a bit out-of-place to mention this here (it would feel "tacked on") and in principle this point should not need to be stated. If, however, we get comments or questions about this (particularly during review), we can revise to add this point later.

\subsection{Simulation details}

Unless otherwise noted, we express all distances in units of $\sigma\equiv\sigma_{AA}$ and all times in units of
\[ \tau_\text{LJ} \equiv \left(m\sigma^2/\epsilon\right)^{1/2}, \]
where $m\equiv m_{AA}$. We performed all simulations using the HOOMD-blue molecular dynamics software package \cite{hoomd-2008, hoomd-2015}, running primarily on GPUs. We compiled HOOMD-blue in single precision and typically used a time step of $\delta t=0.005$. Compiling the software in double precision offered no improvement in the energy drift, owing to our decision to use truncated-and-shifted potentials. HOOMD-blue offers the option of a smooth cutoff that greatly improves the numerical precision, particularly in conjunction with double-precision compilation \cite{joaander-2013}. %% Cite HOOMD paper on precision of methods
However, most of our analysis requires observation times short enough that energy drift is not noticeable using single-precision builds and plain cutoffs. The exception to this is our computation of relaxation times, which requires generating constant-energy trajectories long enough to observe the decay of the self-correlation function. At very low temperatures, the required observation time is very long, and over such long times the energy drift is significant and the self-correlation function is noticeably impacted. In these situations, we use a less aggressive time step of $0.002$, which is enough to ensure that the computed relaxation time is accurate.

In simulations where we need to control the temperature, whether equilibrating or cooling the liquid, we employ a Langevin thermostat in the low-friction limit. %% Cite thermostat [Maybe not necessary]
This choice of thermostat ensures proper sampling when the liquid is in equilibrium, with minimal impact on the dynamics. When the liquid is out of equilibrium, this thermostat also ensures quick relaxation of the kinetic degrees of freedom, such that the velocities may be assumed to follow the Maxwell-Boltzmann distribution at the bath temperature.

\subsection{Calculating non-equilibrium averages}

In this paper, we sometimes refer to averages in and out of equilibrium with some degree of ambiguity. Here, we clarify the meaning of an average and offer some comments about how these averages are computed in practice. To begin, consider a quantity $x$ that is a function of the configuration $\mathcal{C}$ of the system, $x=x(\mathcal{C})$. Of interest is the \emph{ensemble average} of $x$,
\begin{equation}\label{eq:ensemble}
  \left<x\right> \equiv \sum_\mathcal{C} w(\mathcal{C}) x(\mathcal{C}),
\end{equation}
where $w(\mathcal{C})$ is a weight function that determines the contribution of configuration $\mathcal{C}$ to the average. Choosing an ensemble means choosing the function $w$. The micro-canonical ensemble weights all allowed configurations equally, while the canonical ensemble weights configurations by the Boltzmann factor $\exp(-E(\mathcal{C})/T)$, where $E(\mathcal{C})$ is the energy of configuration $\mathcal{C}$ and $T$ is the temperature \cite{imsm}. %% Cite DC's green book
Non-equilibrium ensembles, corresponding to systems driven out of equilibrium by a protocol of time-varying external parameters, are also possible, though an analytical expression for $w$ may be known only in special cases.

In practice, the ensemble average of $x$ is computed by a simple average of $x$ over configurations drawn from the corresponding distribution $w$. If we sample $M$ configurations, then we estimate $\left<x\right>$ with
\begin{equation}\label{eq:average}
  \overline{x} \equiv \frac{1}{M}\sum_{i=1}^M x(\mathcal{C}_i).
\end{equation}
Provided the set of configurations $\left\{\mathcal{C}_i\right\}$ is representative of the ensemble, $\overline{x}$ will be a good estimate of the true ensemble average. For equilibrium ensembles, this representative subset is typically generated by periodically sampling configurations from a sufficiently long time series following a generous period of equilibration.

Non-equilibrium ensembles require specification of an \emph{initial condition}, i.e., the ensemble that prevails at $t=0$, as well as the time dependence of all external parameters for $t>0$. This time dependence means that time averaging is not an option out of equilibrium. Instead, we generate representative subsets by independently preparing a large number of systems according to the protocol, drawing initial configurations from the known ensemble at $t=0$ and then subjecting those configurations to the protocol of time-varying external parameters. The configurations obtained at the end of the protocol comprise a representative subset of the non-equilibrium ensemble. Because the cooling protocols we consider begin with the system in equilibrium at a high initial temperature $T_\text{i}$, the starting configurations are drawn from the canonical ensemble at temperature $T_\text{i}$. All parameters besides the temperature are fixed and $T$ decreases linearly with time, so the rate $\nu_\text{c}$ is sufficient to fully specify a cooling protocol.

So far we have implicitly limited ourselves to cooling protocols that take the temperature all the way down to zero. However, we could imagine cooling to an intermediate temperature $T$ such that $T_\text{i}>T>0$. The rate $\nu_\text{c}$ then corresponds to a family of protocols and associated non-equilibrium ensembles, differentiated by the final temperature $T$ to which the system is cooled. Note that the ensemble corresponding to $\nu_\text{c}$ and $T$ could also be produced by \emph{pausing} the full cooling protocol when it reaches temperature $T$ and sampling the configurations at that point, which is the approach we take in practice. If $x=x(\mathcal{C})$ is our quantity of interest, then we denote the average over this non-equilibrium ensemble by $\left<x(T)\right>_{\nu_\text{c}}$. For a fixed rate $\nu_\text{c}$, this quantity is a function of temperature indicating how the average of $x$ changes as the system is cooled to lower temperatures at rate $\nu_c$. As the rate changes, so does the function.
%, as the relaxation time does in figure \ref{fig:Relaxation_Times}.

For any $\nu_\text{c}>0$, there will be a glass transition temperature, $T_\text{g}$, satisfying equation \ref{eq:define_Tg}, below which this function will deviate substantially from the equilibrium result. In the limit of infinitely slow cooling, $\nu_\text{c}\rightarrow 0$, we expect $\left<x(T)\right>_{\nu_\text{c}}$ to approach the equilibrium ensemble average. In the work that follows, we will typically drop the subscript $\nu_\text{c}$ for ease of notation, but it should be clear from context whether the average is equilibrium or non-equilibrium, and for the latter, what the corresponding rate is.

\subsection{Iso-configurational averaging}

In the preceding discussion of non-equilibrium averages we restricted our attention to static observables that depend on a single configuration. However, in much of the work that follows, we will be concerned with \emph{dynamic} observables that depend on the configuration at multiple points in a trajectory, e.g.,
\[ x = x(\mathcal{C}(t_1),\mathcal{C}(t_2),\dots,\mathcal{C}(t_N)), \]
where $t_1<t_2<\dots<t_N$. Typically we will focus on observables that depend only on the endpoints of the trajectory, so that the above simplifies to
\[ x = x(\mathcal{C}(0),\mathcal{C}(t)), \]
where $t$ is the total length of the trajectory, or observation time.

The so-called \emph{iso-configurational ensemble} \cite{AWC-2004} associated with an initial configuration $\mathcal{C}(0)$, a temperature $T$, and an observation time $t$ is generated by drawing initial velocities from the Maxwell-Boltzmann distribution at temperature $T$ and then integrating Newton's equations of motion forward to time $t$ to produce a new configuration $\mathcal{C}(t)$. The average of a dynamical observable $x$ over this new ensemble is termed the \emph{iso-configurational average}, %% Do we need additional citations for this? Perhaps AWC again, or others that use isoconfigurational averages.
and is itself a static observable that depends only on the initial configuration $\mathcal{C}(0)$. For convenience, we denote this observable by $x_\text{iso}(\mathcal{C})$.

We will often write $\left<x(\mathcal{C}(0),\mathcal{C}(t))\right>$ when we mean to refer to the ensemble average of $x_\text{iso}$, so that the following expressions are equivalent:
\[ \left<x(\mathcal{C}(0),\mathcal{C}(t))\right> \equiv \left<x_\text{iso}\right>. \]
That is, when dealing with a dynamical observable, the angle brackets denote an ensemble average as well as an iso-configurational average. In practice, this average is computed via an extension of equation \ref{eq:average}:
\[ \overline{x} = \frac{1}{M}\sum_{i=1}^M \frac{1}{N_\text{iso}}\sum_{j=1}^{N_\text{iso}} x(\mathcal{C}_i,\mathcal{C}_{ij}(t)). \]
In this expression, $N_\text{iso}$ is the number of trajectories run per initial configuration $\mathcal{C}_i$ to sample its iso-configurational ensemble, and $\mathcal{C}_{ij}(t)$ is the configuration produced at time $t$ in the $j$th trajectory that starts with configuration $\mathcal{C}_i$.

For $\overline{x}$ to produce a good estimate of $\left<x(t)\right>$, it is important that $M$ is large. By contrast, it is \emph{not} particularly important that $N_\text{iso}$ be large, provided $M$ is large. The iso-configurational average is essentially just an average over the initial velocities of the trajectories, and for a sufficiently large value of $M$, this velocity averaging is satisfied even for $N_\text{iso}=1$. In practice, however, the value of $M$ will be limited (by computational resources), and a larger value of $N_\text{iso}$ will sometimes be necessary.

\section{Results and Discussion}

In this section, we present the results of our atomistic simulations and compare them, in detail, to expectations based on the one-dimensional East model. For each model listed in table \ref{tab:params}, we performed cooling protocols as described in Models and Methods, for various cooling rates. For each model and cooling rate, at various values of the external temperature $T$, we compute the following properties of interest:
\begin{enumerate}
  \item[(i)] the relaxation time of the liquid;
  \item[(ii)] the rate at which particles make significant displacements over short observation times;
  \item[(iii)] the spatial distribution of particles that make these short-time displacements.
\end{enumerate}
Each of these properties obeys East model scaling in equilibrium and is expected to deviate from that scaling as $T$ drops below $T_\text{g}$.

In the remainder of this section, we discuss each of these properties in greater detail, our expectations for them (in and out of equilibrium), and how the results of our atomistic simulations compare to these expectations.

\subsection{Relaxation time}

\begin{figure*}[t]
  \centering
  \includegraphics[width=\linewidth]{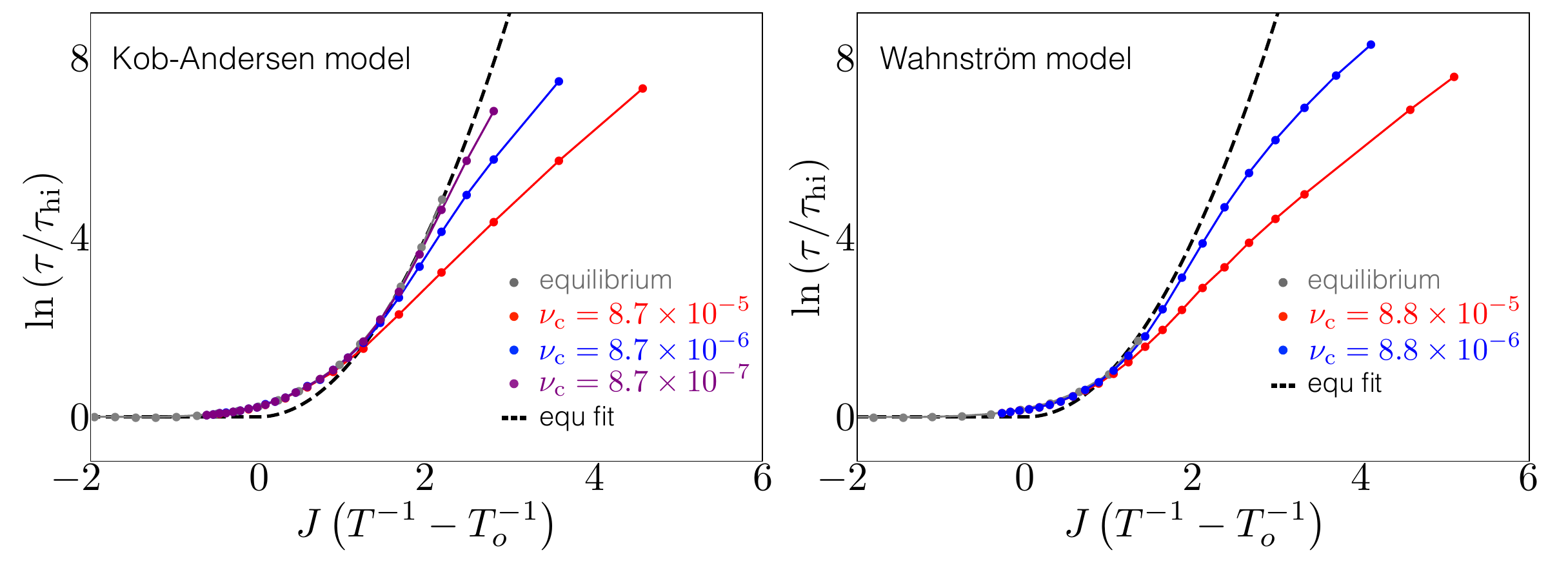}
  \caption{Logarithm of the relaxation time (with the high-temperature Arrhenius contribution removed) versus inverse temperature for the Kob-Andersen (left) and Wahnstr{\"o}m (right) models, for various cooling rates. Rates have units of temperature/time, where time is expressed in units of $\tau_\text{LJ}$. Relaxation times are computed as described in the main text. The dashed black curves show the result of fitting equilibrium data (gray points) to the equilibrium form in equation \ref{eq:tau_equ}. Solid lines are guides to the eye.}
  \label{fig:Relaxation_Times}
\end{figure*}

Following reference \cite{keys-prx-2013}, we define the relaxation time of the liquid as the time $\tau$ satisfying $F_s(q_0,\tau)=1/e$, where
\begin{equation}\label{eq:fskt}
  F_s(k,t) \equiv \left<\exp\left[i\mathbf{k}\cdot\left(\mathbf{r}_i(t)-\mathbf{r}_i(0)\right)\right]\right>
\end{equation}
is the self-correlation function and $q_0$ is the wavevector that maximizes the static structure factor. The angle brackets in equation \ref{eq:fskt} may denote either an equilibrium or non-equilibrium average. In the non-equilibrium case, the average is performed over an ensemble of constant-energy trajectories of length $t$, with initial momenta drawn from the Maxwell-Boltzmann distribution and the initial configuration drawn from a specified non-equilibrium ensemble. See Models and Methods for more detail.

At high temperatures, $T>T_\text{o}$, the relaxation time as defined above is Arrhenius, obeying the linear relationship
\begin{equation}\label{eq:tau_hi_T}
  \ln\tau(T) = \ln\tau_0 + E\left(\frac{1}{T}\right),
\end{equation}
where $\tau_0$ and $E$ are material properties governing the dynamics of the high-temperature liquid. These parameters can easily be determined by fitting equation \ref{eq:tau_hi_T} to relaxation time data computed at high temperatures. The energy scale $E$ may be interpreted as an intrinsic energy barrier that particles must overcome in order to undergo a local reorganization. We expect this energy barrier to be present at all temperatures, so that even for $T<T_\text{o}$, equation \ref{eq:tau_hi_T} contributes to the relaxation time.

As discussed in the Introduction, for $T\ll T_\text{o}$, the relaxation time obeys the parabolic scaling of equation \ref{eq:parabolic_law}. Combining this parabolic contribution with the high-temperature contribution in equation \ref{eq:tau_hi_T} yields
\begin{equation}\label{eq:tau_lo_T}
  \ln\tau(T) = \ln\tau_0 + E\left(\frac{1}{T}\right) + J^2\left(\frac{1}{T}-\frac{1}{T_\text{o}}\right)^2,
\end{equation}
for $T<T_\text{o}$. We can easily combine the high- and low-temperature forms into a single formula valid at all temperatures:
\begin{multline}\label{eq:tau_equ}
  \ln\tau(T) = \ln\tau_\text{o} + E\left(\frac{1}{T}\right) + \\ + J^2\left(\frac{1}{T}-\frac{1}{T_\text{o}}\right)^2\Theta\left(T^{-1}{-}T_\text{o}^{-1}\right),
\end{multline}
where $\Theta(x)$ is the Heaviside step function and ensures that the parabolic contribution only applies at temperatures below $T_\text{o}$.

%Because we are primarily concerned with the behavior of the liquid at $T<T_\text{o}$, it is convenient to consider the relaxation time relative to the high-temperature contribution in equation \ref{eq:tau_hi_T}. To do so, we introduce the notation
%\[ \ln\tau_\text{hi}(T) \equiv \ln\tau_\text{o} + E\left(\frac{1}{T}\right) \]
%and
%\[ \ln\tau_\text{lo}(T) \equiv J^2\left(\frac{1}{T} - \frac{1}{T_\text{o}}\right)^2 \]
%to denote the high- and low-temperature contributions, respectively, to the relaxation time. With this notation, the relaxation time in equation \ref{eq:tau_equ} becomes
%\[ \ln\tau(T) = \ln\tau_\text{hi}(T) + \Theta\left(T^{-1}{-}T_\text{o}^{-1}\right)\ln\tau_\text{lo}(T). \]
%We can then define $\tau_\text{rel}\equiv\tau/\tau_\text{hi}$ to be the relaxation time relative to the high-temperature contribution. It follows that
%\[ \tau_\text{rel}(T) = J^2\left(\frac{1}{T}-\frac{1}{T_\text{o}}\right)^2\Theta\left(T^{-1}{-}T_\text{o}^{-1}\right). \]
%The advantage of working with $\tau_\text{rel}$ is clear: it captures the nontrivial variation of the relaxation time. At high temperatures, $\ln\tau_\text{rel}$ is constant and equal to zero; at low temperatures, it grows parabolically with inverse temperature.

The above equations for the relaxation time are valid for glass-forming liquids in equilibrium. However, as discussed in the Introduction, a liquid subjected to constant-rate cooling will eventually fall out of equilibrium at the glass transition temperature, $T_\text{g}$, given by equation \ref{eq:define_Tg}. At temperatures below $T_\text{g}$, the structure of the liquid is arrested and energetic barriers to relaxation are fixed. Consequently, we expect the relaxation time to exhibit Arrhenius behavior for $T<T_\text{g}$. This super-Arrhenius-to-Arrhenius crossover is frequently observed in experiments and is a key signature of the glass transition \cite{angell-review-2000,Russians-arrhenius-glass,alegria-arrhenius-glass,plazek-magill-i,plazek-magill-iv,arrhenius-glass}. %% Citation needed

For the models listed in table \ref{tab:params}, for each cooling rate considered, we computed the relaxation time $\tau$ at various temperatures $T$. The results are shown in figure \ref{fig:Relaxation_Times}, where we plot $\ln\left(\tau/\tau_\text{hi}\right)$ versus $1/T$ for each combination of model and cooling rate, where $\tau_\text{hi}(T)\equiv\ln\tau_\text{o}+E/T$ denotes the high-temperature contribution in equation \ref{eq:tau_hi_T}. At high temperatures, $T\sim T_\text{o}$, where $\tau$ changes only modestly with $T$, the relaxation time shows little dependence on the cooling rate. However, the results for each cooling rate begin to separate noticeably as temperature is lowered and $\tau$ begins to grow more rapidly with decreasing $T$. The effect of the cooling rate on the relaxation time out of equilibrium is as expected: faster cooling leads $\tau(T)$ to quickly deviate from the equilibrium scaling, whereas slower cooling leads $\tau(T)$ to follow the equilibrium scaling to lower temperatures, consistent with the lower glass transition temperature implied by equation \ref{eq:define_Tg}. Furthermore, for all cooling rates, for both models, $\ln\tau(T)$ versus $1/T$ appears to approach a straight line, consistent with our expectation that the relaxation time exhibits Arrhenius behavior below $T_\text{g}$.

\begin{figure*}[t]
  \centering
  \includegraphics[width=\linewidth]{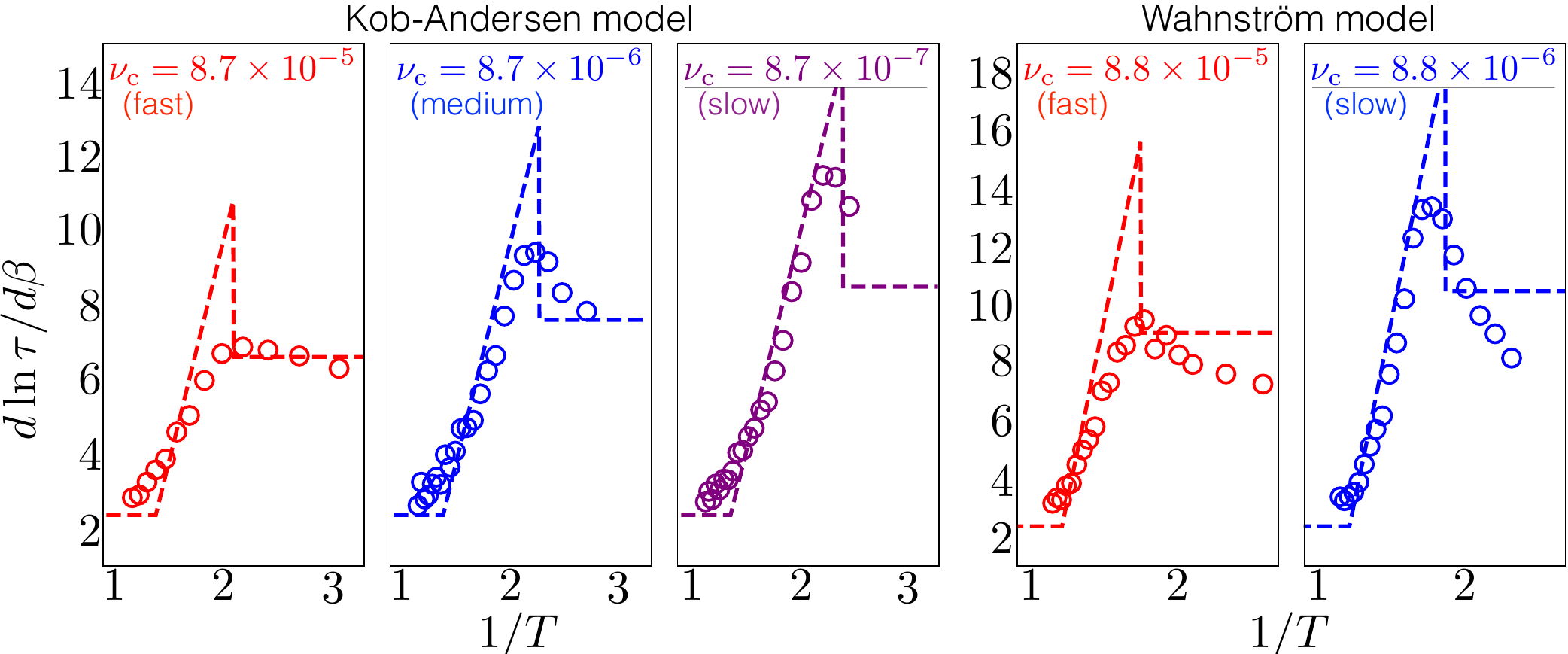}
  \caption{Slope of $\ln\tau(\beta)$, where $\beta=1/T$ is the inverse temperature, for the Kob-Andersen model (left) and the Wahnstr\"{o}m model (right) cooled at varying rates. Open circles show numerical estimates of the slope from the data in figure \ref{fig:Relaxation_Times}, while dashed lines show predictions from East-like scaling, assuming an abrupt glass transition at $T_\text{g}$.}
  \label{fig:Relaxation_Times_Delta}
\end{figure*}

The slope of $\ln\tau(T)$ versus $1/T$ for $T<T_\text{g}$ can be estimated from simple arguments based on the East model. To start, we write the low-temperature contribution to the relaxation time, which we denote $\tau_\text{lo}(T)$, as the product of a (temperature-dependent) free energy barrier and an explicit temperature dependence:
\[ \ln\tau_\text{lo}(T) = F(T)\left(\frac{1}{T}-\frac{1}{T_\text{o}}\right). \]
As discussed in the Introduction, super-Arrhenius relaxation is a consequence of $F(T)$ growing with $1/T$. Arrhenius behavior below $T_\text{g}$, such as that observed in figure \ref{fig:Relaxation_Times}, implies that this barrier has become temperature independent, a consequence of the structural arrest of the liquid as $T$ passes through $T_\text{g}$. Assuming the liquid abruptly falls out of equilibrium at $T_\text{g}$, the relaxation time will obey
\[ \ln\tau_\text{lo}(T) = F(T_\text{g})\left(\frac{1}{T}-\frac{1}{T_\text{o}}\right), \]
for $T<T_\text{g}$. That is, the equilibrium barrier at $T_\text{g}$ is frozen into the material for all $T<T_\text{g}$. %% Additional DL citation here?
In the East model, this barrier is related, through equation \ref{eq:hierarchical}, to the mean inter-excitation spacing, which grows as temperature is lowered according to equation \ref{eq:spacing}. The inter-excitation spacing at the glass transition, denoted $\ell_\text{ne}$, is given by
\[ \ell_\text{ne} = e^{J_0\left(1/T_\text{g}-1/T_\text{o}\right)}. \]
This lengthscale is fixed below $T_\text{g}$ \cite{Limmer-PNAS-2014,Limmer-JCP-2014}, leading to a fixed energy barrier
\begin{align*} F(T_\text{g}) = \gamma J_0\ln\ell_\text{ne} = \gamma J_0^2\left(\frac{1}{T_\text{g}}-\frac{1}{T_\text{o}}\right)
\end{align*}
that prevails below $T_\text{g}$. This barrier is also precisely the out-of-equilibrium slope of $\ln\tau_\text{lo}(T)$ with respect to $1/T$. Defining $J^2\equiv\gamma J_0^2$, the relaxation time for an East-like system with glass transition temperature $T_\text{g}$ is given by
\begin{equation}\label{eq:tau_neq}
  \ln\tau_\text{lo}(T) = 
  \begin{cases}
    J^2{\left(\frac{1}{T}{-}\frac{1}{T_\text{o}}\right)^2}, & T_\text{o}>T>T_\text{g}; \\
    J^2{\left(\frac{1}{T_\text{g}}{-}\frac{1}{T_\text{o}}\right)}{\left(\frac{1}{T}{-}\frac{1}{T_\text{o}}\right)}, & T>T_\text{g}.
  \end{cases}
\end{equation}
From equation \ref{eq:tau_neq}, we can see that the derivative of $\ln\tau_\text{lo}(T)$ with respect to $1/T$ drops discontinuously by a factor of two, from $2J^2\left(1/T_\text{g}-1/T_\text{o}\right)$ immediately before the glass transition to $J^2\left(1/T_\text{g}-1/T_\text{o}\right)$ immediately after. This factor-two change in slope is a corollary of the parabolic law \cite{Limmer-PNAS-2014}. %% This drop in the slope by a factor of one-half is evident in cooling experiments.

Figure \ref{fig:Relaxation_Times_Delta} compares the prediction of the preceding arguments with the results of our atomistic simulations. The dashed lines show the predicted slope, including the discontinuity at $T_\text{g}$. The open circles show numerical estimates of the slope using the data from figure \ref{fig:Relaxation_Times}. For the Kob-Andersen model, for the two slower cooling rates (second and third panes), the computational results show fairly good agreement with the prediction, at least over the limited range of available data. The major difference between our predictions and the simulation results, which is clearly visible for the moderate cooling rate (second pane), is the smoothness of the slope around $T_\text{g}$. However, this is not unexpected; the discontinuity in the predicted slope is a consequence of assuming that the system falls abruptly out of equilibrium at $T_\text{g}$. To the extent that this process is gradual rather than abrupt -- perhaps reflecting the nontrivial distribution of length and energy scales in the system %% Citation
-- the slope will exhibit a peak that is smoothed over a finite temperature range around $T_\text{g}$. This is indeed what we see in figure \ref{fig:Relaxation_Times_Delta}. Interestingly, for the fastest cooling rate, the Kob-Andersen model exhibits almost no peak at all in the slope. This may be a result of cooling too quickly, such that the liquid begins falling out of equilibrium before it clears the crossover region around $T_\text{o}$. Nonetheless, for all three cooling rates, the limiting slope for $T<T_\text{g}$ seems to agree fairly well with the predicted slope, at least within the limits of the data we could obtain with our computational resources.

% Comment about slower cooling protocols producing even better agreement?

The results for the Wahnstr\"{o}m model are considerably worse than for the Kob-Andersen model. Although the predicted values for $T_\text{g}$ seem reasonable, the peaks in the data fall far short of the predicted peaks and, even worse, there is substantial quantitative disagreement in the limiting slope. These discrepancies are likely due to the great difficulty in accurately measuring the parameter $J$ for the Wahnstr\"{o}m model. To obtain a reliable estimate of $J$ requires computing the equilibrium relaxation time at deeply supercooled conditions, far away from the crossover to heterogeneous dynamics at $T_\text{o}$. Unfortunately, the Wahnstr\"{o}m model crystallizes at only modest supercooling, so that only a limited range of data is available to determine this parameter. The Kob-Andersen model resists crystallization to much lower temperatures, so that its $J$ value may be obtained more reliably.

\subsection{Rate of short-time displacements}

\begin{figure*}[t]
  \centering
  \includegraphics[width=\linewidth]{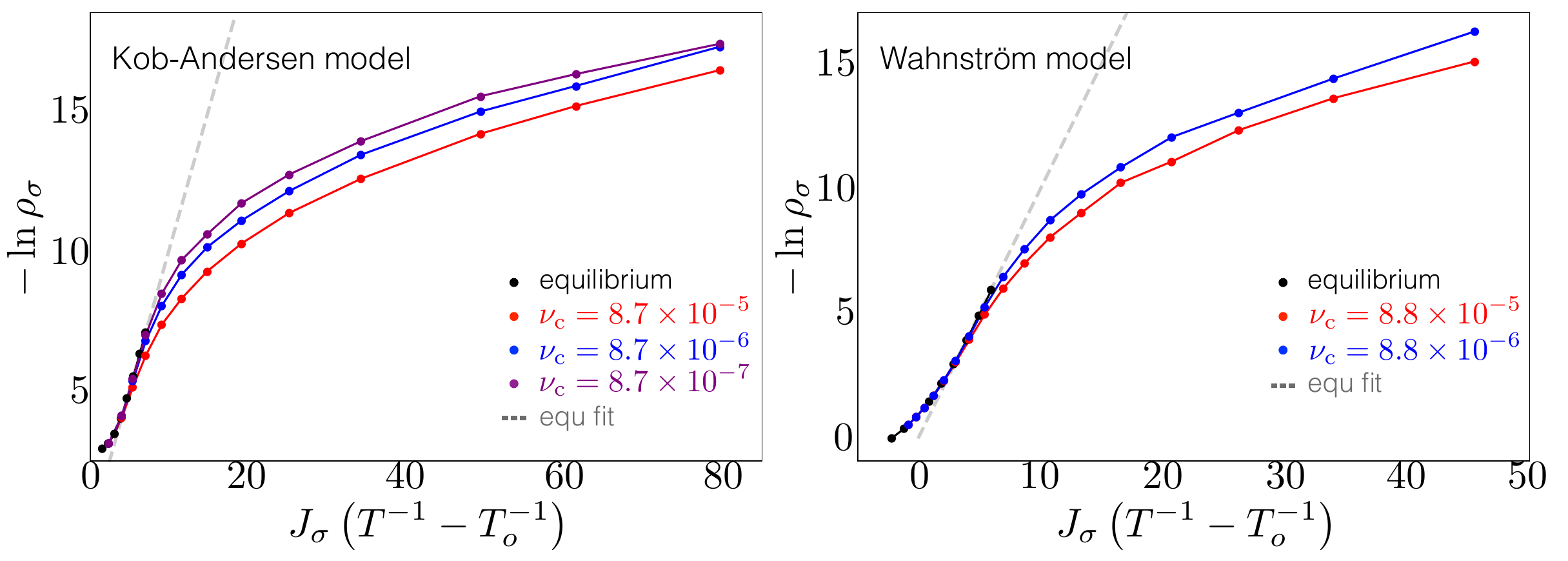}
  \caption{Fraction $\rho_\sigma$ of particles that have displaced at least $\sigma$ in an observation time $t_\text{obs}$. For the Kob-Andersen model (left), we choose $t_\text{obs}=300$. For the Wahnstr\"{o}m model (right), we choose $t_\text{obs}=10$. The shape of the curves changes little for $T<T_\text{g}$ over this range of observation times. The gray dashed lines show the expected equilibrium scaling; solid lines are guides to the eye.}
  \label{fig:Concentration}
\end{figure*}

In supercooled liquids, non-trivial particle dynamics is confined to localized excitations or ``soft spots'' where particles are locally unjammed and can make small (${\sim}\sigma$-sized) displacements over short times \cite{keys-prx-2013}. %% Citation (Keys PRX plus others)
In analogy with the East model, these localized excitations carry a free-energetic cost, which we denote $J_0$, and can facilitate motion in neighboring regions through the creation (and destruction) of neighboring excitations.

Soft spots in supercooled liquids are structural in origin, depending only on the set of particle coordinates. %% Citation
Nonetheless, the construction of a local, static order parameter capable of reliably distinguishing mobile from immobile regions has proven elusive. %% Citation
For this reason, and because non-trivial dynamics occurs only near excitations, we identify soft spots by looking for significant particle displacements over short observation times. We associate particle $i$ with an excitation if, in a short observation time $t$, it displaces a specified minimum distance $a\sim\sigma$. Mathematically, this is accomplished with the following \emph{dynamic} order parameter:
\begin{equation}\label{eq:indicator}
  h_i(t, a) \equiv \Theta\left(\left|\overline{\mathbf{r}}_i(t)-\overline{\mathbf{r}}_i(0)\right| - a\right).
\end{equation}
In the above equation, $\Theta(x)$ is the unit step function, and the overlines above the position vectors indicate inherent structure coordinates. The latter is necessary to remove vibrational motion and can be obtained by steepest descent to the nearest potential energy minimum or by a suitable coarse-graining in time \cite{keys-prx-2013}. %% Cite PRX
If at time $t$ particle $i$ is at least a distance $a$ away from its initial position, then $h_i=1$ and we associate particle $i$ with an excitation.

The order parameter $h_i$ is dynamic, so that it depends on a trajectory rather than a single configuration. Thus, it depends on both the initial positions \emph{and} the initial momenta. To construct an order parameter suitable for detecting excitations (which depend on coordinates and not momenta) requires performing an iso-configurational average over the initial momenta, as discussed in the Methods section. The resulting order parameter, $\left<h_i\right>_\text{iso}$, gives the propensity of particle $i$ to move from the specified initial configuration. Particles with $\left<h_i\right>_\text{iso}\approx 0$ are immobile and not associated with excitations.

In the East model, the concentration or density of excitations decreases according to
\begin{equation}\label{eq:boltzmann}
  c(T) = e^{-J_0/T}
\end{equation}
until the glass transition, at which point the concentration plateaus to a cooling-rate-dependent value. We expect an analogous result to hold for atomistic models. To test this expectation, we consider the quantity
\begin{equation}\label{eq:density}
  \rho_\sigma \equiv \left<\Theta(\left|\overline{\mathbf{r}}_i(t)-\overline{\mathbf{r}}_i(0)\right|-\sigma)\right>,
\end{equation}
which is just the ensemble average of $h_i$ with $a=\sigma$. Note that while the concentration $c$ is a \emph{direct} measurement of the number of excitations in the East model, $\rho_\sigma$ is merely a \emph{proxy} for the number of underlying excitations in an atomistic system. The quantity $\rho_\sigma$ gives the average fraction of particles that move in an observation time $t$. In the East model picture, particle motion occurs only near excitations, so that $\rho_\sigma$ should be proportional to the number of underlying (but unobservable) excitations present in the liquid. To make this concrete, we write
\begin{equation}\label{eq:propto}
  \rho_\sigma(t) = r(t) N_\text{exc},
\end{equation}
where $N_\text{exc}$ is the number of particles in the system associated with an underlying excitation and that could \emph{potentially} move within a short time interval. In practice, only a fraction of the $N_\text{exc}$ particles that can move will do so, such that $\rho_\sigma$ is proportional to the total fraction of particles eligible to move.

We write the proportionality constant $r(t)$ as a function of time to emphasize that it tends to grown with the observation time $t$, as more eligible particles displace. However, the relationship in equation \ref{eq:propto} is meaningful only as long as $t$ is short enough that $N_\text{exc}$, which depends on the instantaneous configuration of the system, may be approximated as constant. For longer times, the system configuration changes appreciably and $\rho_\sigma$ is no longer related to an instantaneous set of ``excited'' or mobile particles. Although this timescale is unknown, it must be at least as long as the mean or typical instanton time $\Delta t$ \cite{keys-prx-2013}, which is the time required for a particle to commit to a new position once it has begun a displacement, and it must be shorter than the structural relaxation time $\tau$. In practice, we typically choose $t\approx\Delta t$. At conditions of deep supercooling, under which the system evolves very slowly and the rate of displacements is low even within excitations (due to the presence of intrinsic energy barriers), we sometimes choose $t\gg\Delta t$ in order to generate a larger set of observations. In all cases, however, we are careful to ensure that the chosen observation time satisfies $t\ll\tau$.

With a suitable analog for the concentration now defined for atomistic systems, we can investigate whether the crossover in $c(T)$ described above for the East model has a counterpart in atomistic systems cooled below the glass transition. Figure \ref{fig:Concentration} shows $\rho_\sigma$ over temperatures ranging from $T\sim T_\text{o}$ to $\beta\gg\beta_\text{g}$ for the set of models and cooling rates considered in the previous section. The results look qualitatively similar to those in figure \ref{fig:Relaxation_Times} for the relaxation time: at high temperatures, the data are identical across cooling rates, but at $T<T_\text{g}$, different cooling rates produce different results.

At high temperatures, the liquid remains in equilibrium and $\ln\rho_\sigma^{-1}$ follows the linear scaling expected from equation \ref{eq:boltzmann}. As $T$ approaches $T_\text{g}$, $\ln\rho_\sigma^{-1}$ begins to deviate from this equilibrium scaling. Consistent with the effect of cooling rate on $T_\text{g}$, slower cooling leads to results that begin their deviation from equilibrium at lower temperatures, so that the equilibrium scaling prevails over a larger temperature range. Additionally, at deep supercooling, $\beta\gg\beta_\text{g}$, $\ln\rho_\sigma^{-1}$ increases as the cooling rate is lowered, implying that slower cooling leads to smaller values of $\rho_\sigma$ and lower density of excitations out of equilibrium. This result is expected from the East model and is also supported by figures \ref{fig:Relaxation_Times} and \ref{fig:Relaxation_Times_Delta}, which show that slower cooling leads to slower relaxation times and larger free-energetic barriers to relaxation.

Also noteworthy in figure \ref{fig:Concentration} is that $\ln\rho_\sigma^{-1}$ slowly approaches linear scaling again at very low temperatures, with the asymptotic slope noticeably smaller than the slope in equilibrium. Interestingly, the out-of-equilibrium slope seems to be independent of the cooling rate; indeed, for $\beta\gg\beta_\text{g}$, the results for each model in figure \ref{fig:Concentration} seem only to differ by their vertical offset. This suggests the existence of an intrinsic energy barrier that controls the rate of short-time displacements \emph{within} excitations, a contribution to the overall rate of displacements $\rho_\sigma$ that would be present in equilibrium but only be clearly manifested at $T<T_\text{g}$, where the number of underlying excitations becomes fixed. Although the East model has no concept of intra-excitation dynamics, it exhibits analogous behavior in the instantaneous rate of spin flips, following different linear scalings above and below $T_\text{g}$. %% Should consider removing this statement, since we don't show the results

\subsection{Spatial distribution of soft spots}

A distinguishing feature of the East model is the emergence of inter-excitation correlations as the lattice is cooled through the glass transition. In reference \cite{keys-pnas-2013}, Keys, Garrahan, and Chandler demonstrate the emergence of correlations between excitations frozen into the lattice by cooling, and relate the cooling-rate-dependent lengthscale of these correlations to the response of the heat capacity upon subsequent heating.

Because the East model Hamiltonian lacks any explicit inter-excitation coupling, excitations should be spatially uncorrelated in equilibrium. However, the East model's nontrivial dynamics can lead to spatial correlations emerging when the lattice is forced out of equilibrium by a time-dependent perturbation such as cooling. Such perturbations impose finite timescales on the system, preventing its return to equilibrium and turning the dynamic correlations introduced by the kinetic constraint into static correlations between excitations.

The nonequilibrium correlation length reported in reference \cite{keys-pnas-2013} follows directly from the hierarchical dynamics of the East model, and we expect that a similar correlation length should emerge between the dynamical indicators of excitations in atomistic models subjected to cooling.

In East model glasses, inter-excitation correlations can be detected by computing $P(\ell)$, the distribution of inter-excitation distances, which has the formal mathematical definition
\begin{equation}\label{eq:define_Pell}
  P(\ell) \equiv \left<n_i n_{i+\ell}\prod_{j=1}^{\ell-1}(1-n_{i+j})\right> \biggr/ \left<n_i\right>,
\end{equation}
where the angle brackets denote, as usual, an ensemble average. The distribution $P(\ell)$ is the probability that a randomly selected excitation is separated by a distance $\ell$ from the closest neighboring excitation. These inter-excitation domains, which lack excitations and are thus inactive, are the physical origin of the barriers to relaxation in the East model. If excitations are energetically uncoupled, as they are in the East model, then it can be easily shown that $P(\ell)$ is exponential, and the distribution of excitations on the lattice is described by a Poisson process in space. In this case, the spatial distribution of excitations is ideal-gas-like and the location of each excitation unrelated to and uncorrelated with the locations of all other excitations.

The exponential distribution of inter-excitation domain sizes that holds at equilibrium gives way to a non-exponential distribution as the lattice is cooled below $T_\text{g}$. This is a non-equilibrium effect that owes to the hierarchical dynamics of the East model: as $T$ decreases, smaller inter-excitation domains -- which relax more quickly -- are annihilated, while larger domains -- which relax more slowly -- persist. Domains longer than a cooling-rate-dependent lengthscale $\ell_\text{ne}$ are unable to relax on the finite timescale of the protocol and will be frozen into the material, while domains shorter than this lengthscale are eliminated. The exponential distribution that prevails at equilibrium thus gives way to a non-monotonic distribution that peaks at $\ell_\text{ne}\neq 0$. Naturally, $\ell_\text{ne}$ increases as the cooling rate decreases, as shown in reference \cite{keys-pnas-2013}.

\begin{figure*}[t]
  \centering
  \includegraphics[]{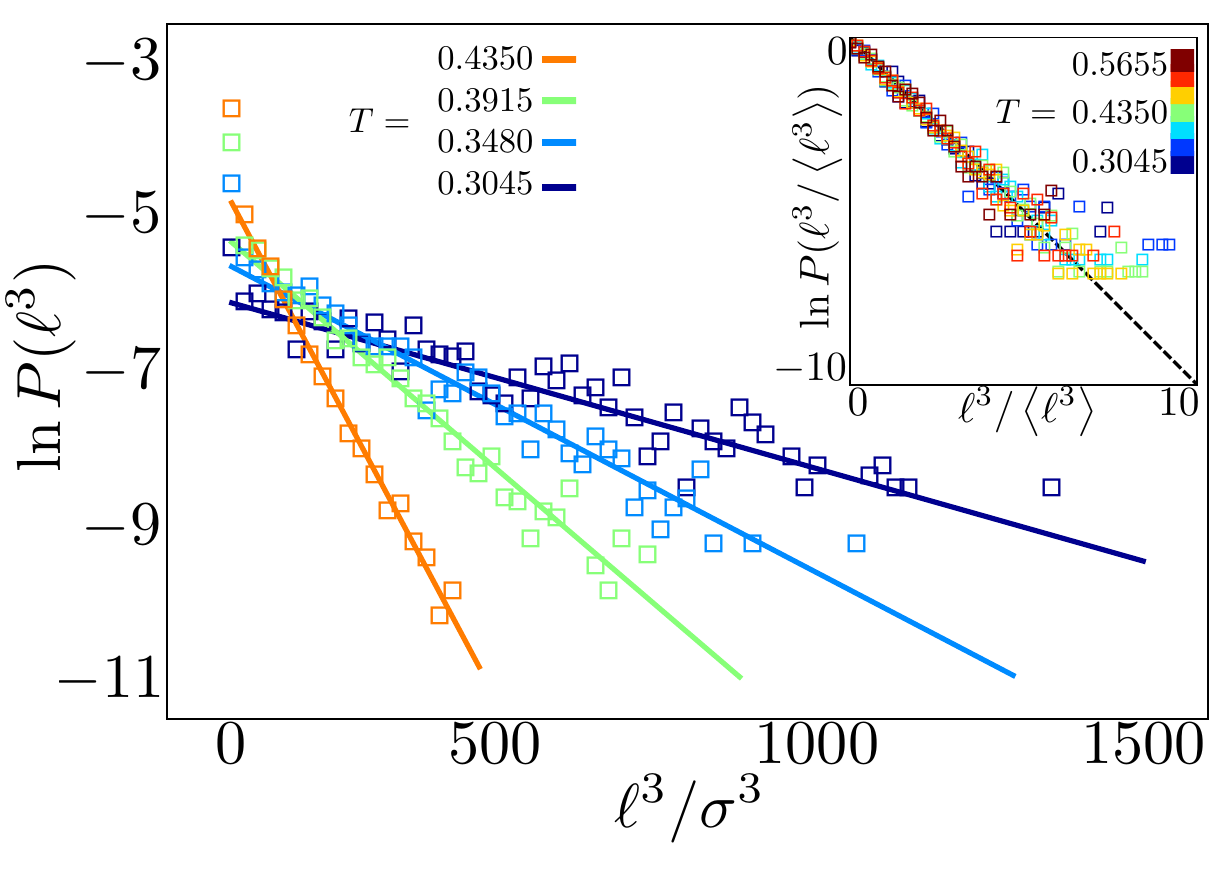}
  \caption{Distribution of $\ell^3$ for the Kob-Andersen model. Results are shown for the slowest cooling protocol that we performed. The main figure shows $P(\ell^3)$ at four temperatures near or below the glass transition temperature $T_\text{g}=0.412$ for this choice of model and cooling rate. For small separations, $\ell\lesssim 3\sigma$, there is a noticeable enhancement in the probability, which we attribute to the collective nature of displacements in a supercooled liquid. Beyond this small-lengthscale enhancement, the distributions are remarkably exponential, as demonstrated by the linearity of the data. The solid curves show linear best fits to the data, excluding the outlier point at $\ell=0$ and any points that represent five or fewer samples. Inset: When the distributions are scaled and shifted by the mean value of $\ell^3$, they collapse onto the exponential form $\ln P(x^3)=-x^3$ (dashed line). For all distributions in this figure, we used the indicator function defined in equation \ref{eq:indicator} with a displacement cutoff $a=0.6\sigma$ and observation time $t_\text{obs}=10$.}
  \label{fig:P_ell}
\end{figure*}

As before, we consider particle displacements, the dynamical indicators of excitations, when working with atomistic systems. To probe the spatial distribution of excitations in atomistic glass formers, we need an analogue of the nearest-neighbor distribution function defined in equation \ref{eq:define_Pell} for the East model. For convenience, we define the quantities
\[ \theta_{ij}(r) \equiv \Theta\left(\left|\mathbf{r}_i(t)-\mathbf{r}_j(0)\right| - r\right) \]
and
\[ \delta_{ij}(r) \equiv \delta\left(\left|\mathbf{r}_i(t)-\mathbf{r}_j(0)\right| - r\right), \]
which are indicator functions that register whether particles $i$ and $j$ are separated at $t=0$ by \emph{at least} the distance $r$ or \emph{exactly} the distance $r$, respectively. With these, we can define
\begin{widetext}
  \begin{equation}\label{eq:define_Pell_atomistic}
    p(\ell) \equiv \left< h_i\prod_{j\neq i} \Big((1 - \theta_{ij}(\ell))(1 - h_j) + \theta_{ij}(\ell) + \delta_{ij}(\ell)h_j \Big)\right> \biggr/ \left<h_i\right>.
  \end{equation}
\end{widetext}
Recall from equation \ref{eq:indicator} that $h_i$ indicates whether particle $i$ has displaced by a minimum distance $a$ in an observation time $t$ and is thus associated with an underlying excitation.

The physical meaning of the distribution $p(\ell)$ defined in equation \ref{eq:define_Pell_atomistic} can be understood by considering the set of all particles that displace a minimum distance $a$, as measured by inherent structure coordinates, over the course of a trajectory (i.e., particles satisfying $h_i=1$). The probability that one of these mobile or ``active'' particles, selected at random, will be separated from the nearest neighboring active particle by a distance $x$, satisfying $\ell\leq x \leq\ell+d\ell$, is given by $p(\ell)d\ell$.

For an ideal gas in $d$ dimensions, it can be shown that this nearest neighbor distribution function has the form
\[ p(\ell) \propto \ell^{d-1}\exp\left(-\left(\ell/\ell_0\right)^d\right), \]
where $\ell_0$ is the average distance between particles in the gas. Although $p(\ell)$ for an ideal gas is exponentially distributed only when $d=1$, a straightforward change of variables from $\ell$ to $\ell^d$ recovers an exponential distribution:
\begin{equation}\label{eq:exp_d}
  p{\left(\ell^d\right)} \propto \exp\left(-\left(\ell/\ell_0\right)^d\right).
\end{equation}
If the excitations in a supercooled liquid are spatially uncorrelated, like the particles in an ideal gas, then they will obey the exponential scaling in equation \ref{eq:exp_d}. It follows that particles displacing over short time intervals, whose positions are strongly correlated with the locations of the excitations, should also obey this scaling. If inter-excitation correlations emerge out of equilibrium, as they do in the East model, then $p{\left(\ell^d\right)}$ should deviate from this scaling.

Figure \ref{fig:P_ell} shows $p{\left(\ell^3\right)}$ at several temperatures $T\lesssim T_\text{g}$ for the Kob-Andersen model subjected to our slowest cooling protocol. For $\ell\gtrsim 3\sigma$, all of the distributions appear remarkably (and unexpectedly) exponential. The only deviation from this exponential scaling is an enhancement of probability at small $\ell$, a feature also observed in references \cite{keys-prx-2013,keys-pre-2015}. This small-lengthscale feature can be understood as manifesting the finite size and localized nature of an excitation, which facilitates dynamics for a collection of particles in a contiguous region of space no more than a few particle diameters in extent. The displacement of a particle signifies the presence of an excitation and, by extension, the proximity of other particles that may readily displace. As a result, particles that displace over short times tend to be highly correlated over short distances.

The preceding observations make clear that $p{\left(\ell^3\right)}$ combines the spatial distribution of underlying excitations with that of displacing particles \emph{within} excitations. As discussed above, the latter contribution is short-ranged and decays on the lengthscale of an excitation, leaving only the inter-excitation distribution at large lengthscales. The distributions plotted in figure \ref{fig:P_ell}, which appear to be exponential for $\ell>3\sigma$, suggest that the spatial distribution of excitations is ideal-gas-like at all temperatures and under all cooling protocols studied. To confirm the exponential scaling at large values of $\ell$, we remove the small-lengthscale ($\ell<3\sigma$) portion of the distributions shown in figure \ref{fig:P_ell}, then shift and normalize what remains. The modified distributions, plotted in the inset of figure \ref{fig:P_ell}, collapse onto the exponential form. Results similar to those in figure \ref{fig:P_ell} were obtained for the Wahnstr\"{o}m model and are presented in the Supplemental Material.

That excitations would be spatially uncorrelated at high temperatures, where the liquid is at or near equilibrium, is expected. That they would \emph{remain} so at temperatures well below $T_\text{g}$ is a surprise given the hierarchical nature of dynamics in supercooled liquids. As discussed previously for the East model, only excitations within $\approx\ell_\text{ne}$ of other excitations can be eliminated in the finite time afforded by the cooling protocol; consequently, the out-of-equilibrium liquid should display negative inter-excitation correlations similar to those observed for the East model \cite{keys-pnas-2013}. The lack of such correlations in our results, at conditions where the liquid is clearly out of equilibrium, is unexpected. To verify the validity of our results, we performed additional calculations, varying the definition of the indicator function, the choices of distance cutoff and observation time, and even the choice of spatial distribution function. The results of many of these calculations are presented in the Supplemental Material. In all cases, at all temperatures, for all models and cooling rates considered, there is no clear evidence of correlations between displacing particles, except for the short-range correlations discussed above.

The most straightforward interpretation of the apparent contradiction between the above result and our prior expectation is that short-time particle displacements are \emph{not} reliable indicators of excitations in atomistic systems, contrary to our stated assumption. This could be the case if many of the observed displacements comprise relatively unimportant (but still non-vibrational) motions that do not contribute to structural relaxation. If so, inter-excitation correlations could be present in the glasses that we prepared, but unobservable using any of the indicator functions that we tested. Direct observation of this phenomenon would require developing indicators capable of faithfully distinguishing between trivial motion unrelated to and uncorrelated with underlying excitations and non-trivial motion facilitated by excitations and responsible for structural relaxation. Whether better indicators or a more complete evaluation of the applicability of the East model is necessary is left to future work.

\section{Conclusions}

The results in the previous section complicate the simple picture of glassy dynamics offered by the one-dimensional East model. Although the out-of-equilibrium behavior of the relaxation time and displacement rate largely agree with our East-model-based expectations, the absence of spatial correlations between displacing particles at temperatures well below $T_\text{g}$
% the lack of correlations in the spatial distribution of displacing particles, for $T<T_\text{g}$,
stood in stark contrast to them. Reconciling this surprising result with our understanding of the East model would represent an important step forward in the study of glassy phenomena and the glassy state.

In the previous section, we suggested the possibility that a failure of our selected indicator functions to faithfully detect underlying excitations may be the source of the discrepancy. If this is the case and a better indicator may be constructed, then the apparent contradiction is easy to resolve and our East-model-based understanding of glassy materials requires at most minor modifications.

However, if more sophisticated indicators confirm the results of this work, then our current picture of glassy phenomena requires more significant revision. Specifically, the lack of spatial correlations between excitations implies that the energy barriers governing relaxation and aging dynamics cannot be solely related to the spatial distribution of an underlying set of excitations, as is the case for the East and other kinetically constrained models. This possibility is supported by the observation in reference \cite{soft-east} that the kinetic stability of vapor-deposited glasses cannot be encoded in the configuration of excitations on the lattice, instead requiring careful tuning of a free parameter.

Given the many successes of the East model in predicting the behavior of glass-forming materials, including in this paper, we suspect that revisions to our theory will preserve East model scaling and similarly feature a hierarchical relaxation mechanism, but include important physical details that clarify the observed absence of inter-excitation correlations below $T_g$.

\acknowledgements{This work was supported by Department of Energy Contract No.\ DE-AC0205CH11231, FWP No.\ CH-PHYS02. We would like to thank Shachi Katira for helpful comments on the manuscript.

% We are deeply indebted to David Chandler, without whom this work would not have been possible. David was intimately involved with this work from the beginning and we regret that it could not be completed before his death. We wish to emphasize that the opinions and perspective expressed in this paper belong solely to the listed authors, AH and KKM. Although David's intellectual contributions to this work warrant his inclusion as an author on this paper, we felt it inappropriate to add his name without his permission. If not for David's untimely passing, he too would have contributed to the writing of the manuscript and likely suggested significant revisions to meet his exacting standards.}

\bibliography{references}

\begin{thebibliography}{35}%
\makeatletter
\providecommand \@ifxundefined [1]{%
 \@ifx{#1\undefined}
}%
\providecommand \@ifnum [1]{%
 \ifnum #1\expandafter \@firstoftwo
 \else \expandafter \@secondoftwo
 \fi
}%
\providecommand \@ifx [1]{%
 \ifx #1\expandafter \@firstoftwo
 \else \expandafter \@secondoftwo
 \fi
}%
\providecommand \natexlab [1]{#1}%
\providecommand \enquote  [1]{``#1''}%
\providecommand \bibnamefont  [1]{#1}%
\providecommand \bibfnamefont [1]{#1}%
\providecommand \citenamefont [1]{#1}%
\providecommand \href@noop [0]{\@secondoftwo}%
\providecommand \href [0]{\begingroup \@sanitize@url \@href}%
\providecommand \@href[1]{\@@startlink{#1}\@@href}%
\providecommand \@@href[1]{\endgroup#1\@@endlink}%
\providecommand \@sanitize@url [0]{\catcode `\\12\catcode `\$12\catcode
  `\&12\catcode `\#12\catcode `\^12\catcode `\_12\catcode `\%12\relax}%
\providecommand \@@startlink[1]{}%
\providecommand \@@endlink[0]{}%
\providecommand \url  [0]{\begingroup\@sanitize@url \@url }%
\providecommand \@url [1]{\endgroup\@href {#1}{\urlprefix }}%
\providecommand \urlprefix  [0]{URL }%
\providecommand \Eprint [0]{\href }%
\providecommand \doibase [0]{http://dx.doi.org/}%
\providecommand \selectlanguage [0]{\@gobble}%
\providecommand \bibinfo  [0]{\@secondoftwo}%
\providecommand \bibfield  [0]{\@secondoftwo}%
\providecommand \translation [1]{[#1]}%
\providecommand \BibitemOpen [0]{}%
\providecommand \bibitemStop [0]{}%
\providecommand \bibitemNoStop [0]{.\EOS\space}%
\providecommand \EOS [0]{\spacefactor3000\relax}%
\providecommand \BibitemShut  [1]{\csname bibitem#1\endcsname}%
\let\auto@bib@innerbib\@empty
%</preamble>
\bibitem [{\citenamefont {Elmatad}\ \emph {et~al.}(2009)\citenamefont
  {Elmatad}, \citenamefont {Chandler},\ and\ \citenamefont
  {Garrahan}}]{corresponding-states-i}%
  \BibitemOpen
  \bibfield  {author} {\bibinfo {author} {\bibfnamefont {Y.~S.}\ \bibnamefont
  {Elmatad}}, \bibinfo {author} {\bibfnamefont {D.}~\bibnamefont {Chandler}}, \
  and\ \bibinfo {author} {\bibfnamefont {J.~P.}\ \bibnamefont {Garrahan}},\
  }\href@noop {} {\bibfield  {journal} {\bibinfo  {journal} {Journal of
  Physical Chemistry B}\ }\textbf {\bibinfo {volume} {113}},\ \bibinfo {pages}
  {5563} (\bibinfo {year} {2009})}\BibitemShut {NoStop}%
\bibitem [{\citenamefont {Elmatad}\ \emph {et~al.}(2010)\citenamefont
  {Elmatad}, \citenamefont {Chandler},\ and\ \citenamefont
  {Garrahan}}]{corresponding-states-ii}%
  \BibitemOpen
  \bibfield  {author} {\bibinfo {author} {\bibfnamefont {Y.~S.}\ \bibnamefont
  {Elmatad}}, \bibinfo {author} {\bibfnamefont {D.}~\bibnamefont {Chandler}}, \
  and\ \bibinfo {author} {\bibfnamefont {J.~P.}\ \bibnamefont {Garrahan}},\
  }\href@noop {} {\bibfield  {journal} {\bibinfo  {journal} {Journal of
  Physical Chemistry B}\ }\textbf {\bibinfo {volume} {114}},\ \bibinfo {pages}
  {17113} (\bibinfo {year} {2010})}\BibitemShut {NoStop}%
\bibitem [{\citenamefont {Ediger}\ \emph {et~al.}(1996)\citenamefont {Ediger},
  \citenamefont {Angell},\ and\ \citenamefont {Nagel}}]{ediger-angell-nagel}%
  \BibitemOpen
  \bibfield  {author} {\bibinfo {author} {\bibfnamefont {M.~D.}\ \bibnamefont
  {Ediger}}, \bibinfo {author} {\bibfnamefont {C.~A.}\ \bibnamefont {Angell}},
  \ and\ \bibinfo {author} {\bibfnamefont {S.~R.}\ \bibnamefont {Nagel}},\
  }\href@noop {} {\bibfield  {journal} {\bibinfo  {journal} {Journal of
  Physical Chemistry}\ }\textbf {\bibinfo {volume} {100}},\ \bibinfo {pages}
  {13200} (\bibinfo {year} {1996})}\BibitemShut {NoStop}%
\bibitem [{\citenamefont {Angell}(1996)}]{angell-review-1996}%
  \BibitemOpen
  \bibfield  {author} {\bibinfo {author} {\bibfnamefont {C.~A.}\ \bibnamefont
  {Angell}},\ }\href@noop {} {\bibfield  {journal} {\bibinfo  {journal}
  {Current Opinion in Solid State and Materials Science}\ }\textbf {\bibinfo
  {volume} {1}},\ \bibinfo {pages} {578} (\bibinfo {year} {1996})}\BibitemShut
  {NoStop}%
\bibitem [{\citenamefont {Angell}\ \emph {et~al.}(2000)\citenamefont {Angell},
  \citenamefont {Ngai}, \citenamefont {McKenna}, \citenamefont {McMillan},\
  and\ \citenamefont {Martin}}]{angell-review-2000}%
  \BibitemOpen
  \bibfield  {author} {\bibinfo {author} {\bibfnamefont {C.~A.}\ \bibnamefont
  {Angell}}, \bibinfo {author} {\bibfnamefont {K.~L.}\ \bibnamefont {Ngai}},
  \bibinfo {author} {\bibfnamefont {G.~B.}\ \bibnamefont {McKenna}}, \bibinfo
  {author} {\bibfnamefont {P.~F.}\ \bibnamefont {McMillan}}, \ and\ \bibinfo
  {author} {\bibfnamefont {S.~W.}\ \bibnamefont {Martin}},\ }\href@noop {}
  {\bibfield  {journal} {\bibinfo  {journal} {Journal of Applied Physics}\
  }\textbf {\bibinfo {volume} {88}},\ \bibinfo {pages} {3113} (\bibinfo {year}
  {2000})}\BibitemShut {NoStop}%
\bibitem [{\citenamefont {Debenedetti}\ and\ \citenamefont
  {Stillinger}(2001)}]{debenedetti-stillinger-review}%
  \BibitemOpen
  \bibfield  {author} {\bibinfo {author} {\bibfnamefont {P.~G.}\ \bibnamefont
  {Debenedetti}}\ and\ \bibinfo {author} {\bibfnamefont {F.~H.}\ \bibnamefont
  {Stillinger}},\ }\href@noop {} {\bibfield  {journal} {\bibinfo  {journal}
  {Nature}\ }\textbf {\bibinfo {volume} {410}},\ \bibinfo {pages} {259}
  (\bibinfo {year} {2001})}\BibitemShut {NoStop}%
\bibitem [{\citenamefont {Biroli}\ and\ \citenamefont
  {Garrahan}(2013)}]{biroli-garrahan-review}%
  \BibitemOpen
  \bibfield  {author} {\bibinfo {author} {\bibfnamefont {G.}~\bibnamefont
  {Biroli}}\ and\ \bibinfo {author} {\bibfnamefont {J.~P.}\ \bibnamefont
  {Garrahan}},\ }\href@noop {} {\bibfield  {journal} {\bibinfo  {journal}
  {Journal of Chemical Physics}\ }\textbf {\bibinfo {volume} {138}},\ \bibinfo
  {pages} {12A301} (\bibinfo {year} {2013})}\BibitemShut {NoStop}%
\bibitem [{\citenamefont {J{\"a}ckle}\ and\ \citenamefont
  {Eisinger}(1991)}]{east-1991}%
  \BibitemOpen
  \bibfield  {author} {\bibinfo {author} {\bibfnamefont {J.}~\bibnamefont
  {J{\"a}ckle}}\ and\ \bibinfo {author} {\bibfnamefont {S.~Z.}\ \bibnamefont
  {Eisinger}},\ }\href@noop {} {\bibfield  {journal} {\bibinfo  {journal}
  {Physik B -- Condensed Matter}\ }\textbf {\bibinfo {volume} {84}},\ \bibinfo
  {pages} {115} (\bibinfo {year} {1991})}\BibitemShut {NoStop}%
\bibitem [{\citenamefont {Sollich}\ and\ \citenamefont
  {Evans}(1999)}]{sollich-evans-1999}%
  \BibitemOpen
  \bibfield  {author} {\bibinfo {author} {\bibfnamefont {P.}~\bibnamefont
  {Sollich}}\ and\ \bibinfo {author} {\bibfnamefont {M.~R.}\ \bibnamefont
  {Evans}},\ }\href@noop {} {\bibfield  {journal} {\bibinfo  {journal}
  {Physical Review Letters}\ }\textbf {\bibinfo {volume} {83}},\ \bibinfo
  {pages} {3238} (\bibinfo {year} {1999})}\BibitemShut {NoStop}%
\bibitem [{\citenamefont {Sollich}\ and\ \citenamefont
  {Evans}(2003)}]{sollich-evans-2003}%
  \BibitemOpen
  \bibfield  {author} {\bibinfo {author} {\bibfnamefont {P.}~\bibnamefont
  {Sollich}}\ and\ \bibinfo {author} {\bibfnamefont {M.~R.}\ \bibnamefont
  {Evans}},\ }\href@noop {} {\bibfield  {journal} {\bibinfo  {journal}
  {Physical Review E}\ }\textbf {\bibinfo {volume} {68}},\ \bibinfo {pages}
  {031504} (\bibinfo {year} {2003})}\BibitemShut {NoStop}%
\bibitem [{\citenamefont {Kim}\ \emph {et~al.}(2017)\citenamefont {Kim},
  \citenamefont {Thorpe}, \citenamefont {Noh}, \citenamefont {Garrahan},
  \citenamefont {Chandler},\ and\ \citenamefont {Jung}}]{dayton-east-model}%
  \BibitemOpen
  \bibfield  {author} {\bibinfo {author} {\bibfnamefont {S.}~\bibnamefont
  {Kim}}, \bibinfo {author} {\bibfnamefont {D.~G.}\ \bibnamefont {Thorpe}},
  \bibinfo {author} {\bibfnamefont {C.}~\bibnamefont {Noh}}, \bibinfo {author}
  {\bibfnamefont {J.~P.}\ \bibnamefont {Garrahan}}, \bibinfo {author}
  {\bibfnamefont {D.}~\bibnamefont {Chandler}}, \ and\ \bibinfo {author}
  {\bibfnamefont {Y.}~\bibnamefont {Jung}},\ }\href@noop {} {\bibfield
  {journal} {\bibinfo  {journal} {Journal of Chemical Physics}\ }\textbf
  {\bibinfo {volume} {147}},\ \bibinfo {pages} {084504} (\bibinfo {year}
  {2017})}\BibitemShut {NoStop}%
\bibitem [{\citenamefont {Aldous}\ and\ \citenamefont
  {Diaconis}(2002)}]{aldous-east-model-2002}%
  \BibitemOpen
  \bibfield  {author} {\bibinfo {author} {\bibfnamefont {D.}~\bibnamefont
  {Aldous}}\ and\ \bibinfo {author} {\bibfnamefont {P.}~\bibnamefont
  {Diaconis}},\ }\href@noop {} {\bibfield  {journal} {\bibinfo  {journal}
  {Journal of Statistical Physics}\ }\textbf {\bibinfo {volume} {107}},\
  \bibinfo {pages} {945} (\bibinfo {year} {2002})}\BibitemShut {NoStop}%
\bibitem [{\citenamefont {Chleboun}\ \emph {et~al.}(2013)\citenamefont
  {Chleboun}, \citenamefont {Faggionato},\ and\ \citenamefont
  {Martinelli}}]{chleboun-east-model-2013}%
  \BibitemOpen
  \bibfield  {author} {\bibinfo {author} {\bibfnamefont {P.}~\bibnamefont
  {Chleboun}}, \bibinfo {author} {\bibfnamefont {A.}~\bibnamefont
  {Faggionato}}, \ and\ \bibinfo {author} {\bibfnamefont {F.}~\bibnamefont
  {Martinelli}},\ }\href@noop {} {\bibfield  {journal} {\bibinfo  {journal}
  {Journal of Statistical Mechanics: Theory and Experiment}\ }\textbf {\bibinfo
  {volume} {2013}},\ \bibinfo {pages} {L04001} (\bibinfo {year}
  {2013})}\BibitemShut {NoStop}%
\bibitem [{\citenamefont {Keys}\ \emph {et~al.}(2011)\citenamefont {Keys},
  \citenamefont {Hedges}, \citenamefont {Garrahan}, \citenamefont {Glotzer},\
  and\ \citenamefont {Chandler}}]{keys-prx-2013}%
  \BibitemOpen
  \bibfield  {author} {\bibinfo {author} {\bibfnamefont {A.~S.}\ \bibnamefont
  {Keys}}, \bibinfo {author} {\bibfnamefont {L.~O.}\ \bibnamefont {Hedges}},
  \bibinfo {author} {\bibfnamefont {J.~P.}\ \bibnamefont {Garrahan}}, \bibinfo
  {author} {\bibfnamefont {S.~C.}\ \bibnamefont {Glotzer}}, \ and\ \bibinfo
  {author} {\bibfnamefont {D.}~\bibnamefont {Chandler}},\ }\href@noop {}
  {\bibfield  {journal} {\bibinfo  {journal} {Physical Review X}\ }\textbf
  {\bibinfo {volume} {1}},\ \bibinfo {pages} {021013} (\bibinfo {year}
  {2011})}\BibitemShut {NoStop}%
\bibitem [{\citenamefont {Limmer}\ and\ \citenamefont
  {Chandler}(2014)}]{Limmer-PNAS-2014}%
  \BibitemOpen
  \bibfield  {author} {\bibinfo {author} {\bibfnamefont {D.~T.}\ \bibnamefont
  {Limmer}}\ and\ \bibinfo {author} {\bibfnamefont {D.}~\bibnamefont
  {Chandler}},\ }\href@noop {} {\bibfield  {journal} {\bibinfo  {journal}
  {Proceedings of the National Academy of Sciences}\ }\textbf {\bibinfo
  {volume} {111}},\ \bibinfo {pages} {9413} (\bibinfo {year}
  {2014})}\BibitemShut {NoStop}%
\bibitem [{\citenamefont {Limmer}(2014)}]{Limmer-JCP-2014}%
  \BibitemOpen
  \bibfield  {author} {\bibinfo {author} {\bibfnamefont {D.~T.}\ \bibnamefont
  {Limmer}},\ }\href@noop {} {\bibfield  {journal} {\bibinfo  {journal}
  {Journal of Chemical Physics}\ }\textbf {\bibinfo {volume} {140}},\ \bibinfo
  {pages} {214509} (\bibinfo {year} {2014})}\BibitemShut {NoStop}%
\bibitem [{\citenamefont {Wahnstr{\"o}m}(1991)}]{Wahnstrom-1991}%
  \BibitemOpen
  \bibfield  {author} {\bibinfo {author} {\bibfnamefont {G.}~\bibnamefont
  {Wahnstr{\"o}m}},\ }\href@noop {} {\bibfield  {journal} {\bibinfo  {journal}
  {Physical Review A}\ }\textbf {\bibinfo {volume} {44}},\ \bibinfo {pages}
  {3752} (\bibinfo {year} {1991})}\BibitemShut {NoStop}%
\bibitem [{\citenamefont {Kob}\ and\ \citenamefont
  {Andersen}(1995)}]{Kob-Andersen-1995}%
  \BibitemOpen
  \bibfield  {author} {\bibinfo {author} {\bibfnamefont {W.}~\bibnamefont
  {Kob}}\ and\ \bibinfo {author} {\bibfnamefont {H.~C.}\ \bibnamefont
  {Andersen}},\ }\href@noop {} {\bibfield  {journal} {\bibinfo  {journal}
  {Physical Review E}\ }\textbf {\bibinfo {volume} {51}},\ \bibinfo {pages}
  {4626} (\bibinfo {year} {1995})}\BibitemShut {NoStop}%
\bibitem [{\citenamefont {Keys}\ \emph {et~al.}(2013)\citenamefont {Keys},
  \citenamefont {Garrahan},\ and\ \citenamefont {Chandler}}]{keys-pnas-2013}%
  \BibitemOpen
  \bibfield  {author} {\bibinfo {author} {\bibfnamefont {A.~S.}\ \bibnamefont
  {Keys}}, \bibinfo {author} {\bibfnamefont {J.~P.}\ \bibnamefont {Garrahan}},
  \ and\ \bibinfo {author} {\bibfnamefont {D.}~\bibnamefont {Chandler}},\
  }\href@noop {} {\bibfield  {journal} {\bibinfo  {journal} {Proceedings of the
  National Academy of Sciences}\ }\textbf {\bibinfo {volume} {110}},\ \bibinfo
  {pages} {4482} (\bibinfo {year} {2013})}\BibitemShut {NoStop}%
\bibitem [{\citenamefont {Keys}\ \emph {et~al.}(2015)\citenamefont {Keys},
  \citenamefont {Chandler},\ and\ \citenamefont {Garrahan}}]{keys-pre-2015}%
  \BibitemOpen
  \bibfield  {author} {\bibinfo {author} {\bibfnamefont {A.~S.}\ \bibnamefont
  {Keys}}, \bibinfo {author} {\bibfnamefont {D.}~\bibnamefont {Chandler}}, \
  and\ \bibinfo {author} {\bibfnamefont {J.~P.}\ \bibnamefont {Garrahan}},\
  }\href@noop {} {\bibfield  {journal} {\bibinfo  {journal} {Physical Review
  E}\ }\textbf {\bibinfo {volume} {92}},\ \bibinfo {pages} {022304} (\bibinfo
  {year} {2015})}\BibitemShut {NoStop}%
\bibitem [{\citenamefont {Toxvaerd}\ \emph {et~al.}(2009)\citenamefont
  {Toxvaerd}, \citenamefont {Pedersen}, \citenamefont {Schr{\o}der},\ and\
  \citenamefont {Dyre}}]{Pedersen-JCP-2009}%
  \BibitemOpen
  \bibfield  {author} {\bibinfo {author} {\bibfnamefont {S.}~\bibnamefont
  {Toxvaerd}}, \bibinfo {author} {\bibfnamefont {U.~R.}\ \bibnamefont
  {Pedersen}}, \bibinfo {author} {\bibfnamefont {T.~B.}\ \bibnamefont
  {Schr{\o}der}}, \ and\ \bibinfo {author} {\bibfnamefont {J.~C.}\ \bibnamefont
  {Dyre}},\ }\href@noop {} {\bibfield  {journal} {\bibinfo  {journal} {Journal
  of Chemical Physics}\ }\textbf {\bibinfo {volume} {130}},\ \bibinfo {pages}
  {224501} (\bibinfo {year} {2009})}\BibitemShut {NoStop}%
\bibitem [{\citenamefont {Pedersen}\ \emph {et~al.}(2010)\citenamefont
  {Pedersen}, \citenamefont {Schr{\o}der}, \citenamefont {Dyre},\ and\
  \citenamefont {Harrowell}}]{Pedersen-PRL-2010}%
  \BibitemOpen
  \bibfield  {author} {\bibinfo {author} {\bibfnamefont {U.~R.}\ \bibnamefont
  {Pedersen}}, \bibinfo {author} {\bibfnamefont {T.~B.}\ \bibnamefont
  {Schr{\o}der}}, \bibinfo {author} {\bibfnamefont {J.~C.}\ \bibnamefont
  {Dyre}}, \ and\ \bibinfo {author} {\bibfnamefont {P.}~\bibnamefont
  {Harrowell}},\ }\href@noop {} {\bibfield  {journal} {\bibinfo  {journal}
  {Physical Review Letters}\ }\textbf {\bibinfo {volume} {104}},\ \bibinfo
  {pages} {105701} (\bibinfo {year} {2010})}\BibitemShut {NoStop}%
\bibitem [{\citenamefont {Pedersen}\ \emph {et~al.}()\citenamefont {Pedersen},
  \citenamefont {Bailey}, \citenamefont {Dyre},\ and\ \citenamefont
  {Schr{\o}der}}]{Pedersen-arxiv-2007}%
  \BibitemOpen
  \bibfield  {author} {\bibinfo {author} {\bibfnamefont {U.~R.}\ \bibnamefont
  {Pedersen}}, \bibinfo {author} {\bibfnamefont {N.~P.}\ \bibnamefont
  {Bailey}}, \bibinfo {author} {\bibfnamefont {J.~C.}\ \bibnamefont {Dyre}}, \
  and\ \bibinfo {author} {\bibfnamefont {T.~B.}\ \bibnamefont {Schr{\o}der}},\
  }\href@noop {} {\ }\Eprint {http://arxiv.org/abs/arXiv:0706.0813}
  {arXiv:0706.0813} \BibitemShut {NoStop}%
\bibitem [{\citenamefont {Ninarello}\ \emph {et~al.}(2017)\citenamefont
  {Ninarello}, \citenamefont {Berthier},\ and\ \citenamefont
  {Coslovich}}]{Berthier-2017-models}%
  \BibitemOpen
  \bibfield  {author} {\bibinfo {author} {\bibfnamefont {A.}~\bibnamefont
  {Ninarello}}, \bibinfo {author} {\bibfnamefont {L.}~\bibnamefont {Berthier}},
  \ and\ \bibinfo {author} {\bibfnamefont {D.}~\bibnamefont {Coslovich}},\
  }\href@noop {} {\bibfield  {journal} {\bibinfo  {journal} {Physical Review
  X}\ }\textbf {\bibinfo {volume} {7}},\ \bibinfo {pages} {021039} (\bibinfo
  {year} {2017})}\BibitemShut {NoStop}%
\bibitem [{\citenamefont {Anderson}\ \emph {et~al.}(2008)\citenamefont
  {Anderson}, \citenamefont {Lorenz},\ and\ \citenamefont
  {Travesset}}]{hoomd-2008}%
  \BibitemOpen
  \bibfield  {author} {\bibinfo {author} {\bibfnamefont {J.~A.}\ \bibnamefont
  {Anderson}}, \bibinfo {author} {\bibfnamefont {C.~D.}\ \bibnamefont
  {Lorenz}}, \ and\ \bibinfo {author} {\bibfnamefont {A.}~\bibnamefont
  {Travesset}},\ }\href@noop {} {\bibfield  {journal} {\bibinfo  {journal}
  {Journal of Computational Physics}\ }\textbf {\bibinfo {volume} {227}},\
  \bibinfo {pages} {5342} (\bibinfo {year} {2008})}\BibitemShut {NoStop}%
\bibitem [{\citenamefont {Glaser}\ \emph {et~al.}(2015)\citenamefont {Glaser},
  \citenamefont {Nguyen}, \citenamefont {Anderson}, \citenamefont {Lui},
  \citenamefont {Spiga}, \citenamefont {Millan}, \citenamefont {Morse},\ and\
  \citenamefont {Glotzer}}]{hoomd-2015}%
  \BibitemOpen
  \bibfield  {author} {\bibinfo {author} {\bibfnamefont {J.}~\bibnamefont
  {Glaser}}, \bibinfo {author} {\bibfnamefont {T.~D.}\ \bibnamefont {Nguyen}},
  \bibinfo {author} {\bibfnamefont {J.~A.}\ \bibnamefont {Anderson}}, \bibinfo
  {author} {\bibfnamefont {P.}~\bibnamefont {Lui}}, \bibinfo {author}
  {\bibfnamefont {F.}~\bibnamefont {Spiga}}, \bibinfo {author} {\bibfnamefont
  {J.~A.}\ \bibnamefont {Millan}}, \bibinfo {author} {\bibfnamefont {D.~C.}\
  \bibnamefont {Morse}}, \ and\ \bibinfo {author} {\bibfnamefont {S.~C.}\
  \bibnamefont {Glotzer}},\ }\href@noop {} {\bibfield  {journal} {\bibinfo
  {journal} {Computer Physics Communications}\ }\textbf {\bibinfo {volume}
  {192}},\ \bibinfo {pages} {97} (\bibinfo {year} {2015})}\BibitemShut
  {NoStop}%
\bibitem [{\citenamefont {Anderson}\ and\ \citenamefont
  {Glotzer}()}]{joaander-2013}%
  \BibitemOpen
  \bibfield  {author} {\bibinfo {author} {\bibfnamefont {J.~A.}\ \bibnamefont
  {Anderson}}\ and\ \bibinfo {author} {\bibfnamefont {S.~C.}\ \bibnamefont
  {Glotzer}},\ }\href@noop {} {\ }\Eprint
  {http://arxiv.org/abs/arXiv:1308.5587} {arXiv:1308.5587} \BibitemShut
  {NoStop}%
\bibitem [{\citenamefont {Chandler}(1987)}]{imsm}%
  \BibitemOpen
  \bibfield  {author} {\bibinfo {author} {\bibfnamefont {D.}~\bibnamefont
  {Chandler}},\ }\href@noop {} {\emph {\bibinfo {title} {Introduction to Modern
  Statistical Mechanics}}}\ (\bibinfo  {publisher} {Oxford University Press},\
  \bibinfo {year} {1987})\BibitemShut {NoStop}%
\bibitem [{\citenamefont {Widmer-Cooper}\ \emph {et~al.}(2004)\citenamefont
  {Widmer-Cooper}, \citenamefont {Harrowell},\ and\ \citenamefont
  {Fynewever}}]{AWC-2004}%
  \BibitemOpen
  \bibfield  {author} {\bibinfo {author} {\bibfnamefont {A.}~\bibnamefont
  {Widmer-Cooper}}, \bibinfo {author} {\bibfnamefont {P.}~\bibnamefont
  {Harrowell}}, \ and\ \bibinfo {author} {\bibfnamefont {H.}~\bibnamefont
  {Fynewever}},\ }\href@noop {} {\bibfield  {journal} {\bibinfo  {journal}
  {Physical Review Letters}\ }\textbf {\bibinfo {volume} {93}},\ \bibinfo
  {pages} {135701} (\bibinfo {year} {2004})}\BibitemShut {NoStop}%
\bibitem [{\citenamefont {Mazurin}\ \emph {et~al.}(1982)\citenamefont
  {Mazurin}, \citenamefont {Startsev},\ and\ \citenamefont
  {Stoljar}}]{Russians-arrhenius-glass}%
  \BibitemOpen
  \bibfield  {author} {\bibinfo {author} {\bibfnamefont {O.~V.}\ \bibnamefont
  {Mazurin}}, \bibinfo {author} {\bibfnamefont {Y.~K.}\ \bibnamefont
  {Startsev}}, \ and\ \bibinfo {author} {\bibfnamefont {S.~V.}\ \bibnamefont
  {Stoljar}},\ }\href@noop {} {\bibfield  {journal} {\bibinfo  {journal}
  {Journal of Non-Crystalline Solids}\ }\textbf {\bibinfo {volume} {52}},\
  \bibinfo {pages} {105} (\bibinfo {year} {1982})}\BibitemShut {NoStop}%
\bibitem [{\citenamefont {Alegr\'ia}\ \emph {et~al.}(1995)\citenamefont
  {Alegr\'ia}, \citenamefont {Guerrica-Echevarr\'ia}, \citenamefont
  {Goitiand\'ia}, \citenamefont {Teller\'ia},\ and\ \citenamefont
  {Colmenero}}]{alegria-arrhenius-glass}%
  \BibitemOpen
  \bibfield  {author} {\bibinfo {author} {\bibfnamefont {A.}~\bibnamefont
  {Alegr\'ia}}, \bibinfo {author} {\bibfnamefont {E.}~\bibnamefont
  {Guerrica-Echevarr\'ia}}, \bibinfo {author} {\bibfnamefont {L.}~\bibnamefont
  {Goitiand\'ia}}, \bibinfo {author} {\bibfnamefont {I.}~\bibnamefont
  {Teller\'ia}}, \ and\ \bibinfo {author} {\bibfnamefont {J.}~\bibnamefont
  {Colmenero}},\ }\href@noop {} {\bibfield  {journal} {\bibinfo  {journal}
  {Macromolecules}\ }\textbf {\bibinfo {volume} {28}},\ \bibinfo {pages} {1516}
  (\bibinfo {year} {1995})}\BibitemShut {NoStop}%
\bibitem [{\citenamefont {Plazek}\ and\ \citenamefont
  {Magill}(1966)}]{plazek-magill-i}%
  \BibitemOpen
  \bibfield  {author} {\bibinfo {author} {\bibfnamefont {D.~J.}\ \bibnamefont
  {Plazek}}\ and\ \bibinfo {author} {\bibfnamefont {J.~H.}\ \bibnamefont
  {Magill}},\ }\href@noop {} {\bibfield  {journal} {\bibinfo  {journal}
  {Journal of Chemical Physics}\ }\textbf {\bibinfo {volume} {45}},\ \bibinfo
  {pages} {3038} (\bibinfo {year} {1966})}\BibitemShut {NoStop}%
\bibitem [{\citenamefont {Plazek}\ and\ \citenamefont
  {Magill}(1968)}]{plazek-magill-iv}%
  \BibitemOpen
  \bibfield  {author} {\bibinfo {author} {\bibfnamefont {D.~J.}\ \bibnamefont
  {Plazek}}\ and\ \bibinfo {author} {\bibfnamefont {J.~H.}\ \bibnamefont
  {Magill}},\ }\href@noop {} {\bibfield  {journal} {\bibinfo  {journal}
  {Journal of Chemical Physics}\ }\textbf {\bibinfo {volume} {49}},\ \bibinfo
  {pages} {3678} (\bibinfo {year} {1968})}\BibitemShut {NoStop}%
\bibitem [{\citenamefont {Wojnarowska}\ \emph {et~al.}(2014)\citenamefont
  {Wojnarowska}, \citenamefont {Ngai},\ and\ \citenamefont
  {Paluch}}]{arrhenius-glass}%
  \BibitemOpen
  \bibfield  {author} {\bibinfo {author} {\bibfnamefont {Z.}~\bibnamefont
  {Wojnarowska}}, \bibinfo {author} {\bibfnamefont {K.~L.}\ \bibnamefont
  {Ngai}}, \ and\ \bibinfo {author} {\bibfnamefont {M.}~\bibnamefont
  {Paluch}},\ }\href@noop {} {\bibfield  {journal} {\bibinfo  {journal} {The
  Journal of Chemical Physics}\ }\textbf {\bibinfo {volume} {140}},\ \bibinfo
  {pages} {174502} (\bibinfo {year} {2014})}\BibitemShut {NoStop}%
\bibitem [{\citenamefont {Guti\'errez}\ and\ \citenamefont
  {Garrahan}(2016)}]{soft-east}%
  \BibitemOpen
  \bibfield  {author} {\bibinfo {author} {\bibfnamefont {R.}~\bibnamefont
  {Guti\'errez}}\ and\ \bibinfo {author} {\bibfnamefont {J.~P.}\ \bibnamefont
  {Garrahan}},\ }\href@noop {} {\bibfield  {journal} {\bibinfo  {journal}
  {Journal of Statistical Mechanics: Theory and Experiment}\ }\textbf {\bibinfo
  {volume} {2016}},\ \bibinfo {pages} {074005} (\bibinfo {year}
  {2016})}\BibitemShut {NoStop}%
\end{thebibliography}%

\end{document}